\documentclass[aps,pra,floatfix,superscriptaddress,twocolumn,showpacs,showkeys,10pt]{revtex4-1}
\usepackage{amssymb, amsmath,color,mciteplus,graphicx,subfigure}
\usepackage{calc,epsfig,epstopdf,color,mciteplus,bm,mathrsfs}
\usepackage{times}
\usepackage{float}
\usepackage{lipsum}

\begin{document}

\title{Universal robust nonadiabatic geometric quantum computation with shortened composite pulses on superconducting circuits}

\author{Cheng-Yun Ding}\email{cyding@aqnu.edu.cn}
\affiliation{School of Physics and Astronomy, Anqing Normal University, Anqing 246133, China}

\author{Na Kuang}
\affiliation{School of Physics and Astronomy, Anqing Normal University, Anqing 246133, China}

\author{Yu Jin}
\affiliation{School of Physics and Astronomy, Anqing Normal University, Anqing 246133, China}

\author{Li Tang}
\affiliation{School of Physics and Astronomy, Anqing Normal University, Anqing 246133, China}

\author{Rui Huang}
\affiliation{School of Physics and Astronomy, Anqing Normal University, Anqing 246133, China}

\author{Li-Hua Zhang}
\affiliation{School of Electronic Information and Integrated Circuits, Anqing Normal University,
Anqing 246133, China}

\date{\today}

\begin{abstract}
The nonadiabatic geometric quantum computation (NGQC) using the nonadiabatic geometric phases is a promising strategy to realize the high-fidelity quantum operation for large-scale quantum processing. However, conventional NQGC schemes combining with optimal control techniques to further improve the gate fidelity and robustness usually either comes at the cost of prolonged gate time or require complex pulse control and only are limited to several certain specific quantum gates or/and for a certain type error source. Thus, implementing universal and robust geometric gates with simple control while further suppressing gate times remains a major challenge. Here, we propose a scheme for universal robust geometric gates with shortened composite pulses, where the pulse control is simple and enough available evolution parameters can be utilized to effectively reduce the sensitivity to universal systematic errors. Numerical simulations demonstrate that our universal geometric gates set exhibit much stronger error resilience than their conventional single-loop counterparts. Finally, we also verify the robustness advantage of our scheme on realistic superconducting quantum circuits with experimentally demonstrated technology. Therefore, our work paves an alternative way to achieve high-fidelity and robust geometric quantum computation against multiple errors.

\end{abstract}

\maketitle
\section{Introduction}
Unlike classical computation, quantum computation, based on the nature of quantum coherence and entanglement, can effectively solve several hard problems \cite{shor1997polynomial,PhysRevLett.79.325} that classical computation cannot. To realize a quantum computer, a set of universal quantum logic gates serves as the fundamental building block \cite{PhysRevLett.75.346,PhysRevA.52.3457}. However, in practical experimental implementation, decoherence and control errors will lead to the loss of gate fidelity, resulting in computational errors and failure. Therefore, high-fidelity and robust quantum control is one of the key elements for implementing large-scale quantum computation.

As we all know, geometric quantum control using the geometric phases is believed to be one of promising way to realize robust quantum gates against local systematic errors \cite{PhysRevLett.91.090404,PhysRevA.72.020301,PhysRevLett.102.030404}, as the geometric phase has global characteristics which only depends on the area of evolution path instead of the evolution details. The earliest geometric phase is Berry phase \cite{berry1984quantal} based on adiabatic and cyclic evolution, and thus the constructed quantum gates were easily influenced by decoherence-induced errors due to slow adiabatic evolution, which hinders the use of GQC in real experiments \cite{jones2000geometric,PhysRevA.87.032326,PhysRevLett.122.010503}. Later, Aharonov and Anandan removed the adiabatic limitation of the Berry phase, and proposed the nonabiabatic geometric phase \cite{PhysRevLett.58.1593}. Then, nonadiabatic GQC (NGQC) schemes based the nonabiabatic geometric phase were proposed \cite{PhysRevLett.87.097901,PhysRevLett.89.097902,PhysRevLett.91.187902} and verified experimentally \cite{leibfried2003experimental,PhysRevA.74.020302,song2017continuous}, which can greatly speed up the geometric gates.

Currently, many different NGQC protocols have been developed \cite{PhysRevA.96.052316,PhysRevApplied.10.054051,PhysRevA.101.052302,PhysRevA.103.032609}; however, after experimental verification in various quantum systems \cite{PhysRevLett.124.230503,zhao2021experimental,zhou2025experimental}, the errors and the itself limitations of protocols make the advantages of geometric phase are compromised. To this end, many protocols optimize gate time and/or gate robustness by combining various optimal control technologies, such as composite pulses \cite{PhysRevApplied.10.054051,PhysRevA.103.032609,PhysRevA.109.042615,ding2025composite}, time optimal control \cite{PhysRevApplied.14.064009,PhysRevA.108.042617}, dynamical decoupling \cite{PhysRevA.90.022323,PhysRevA.102.032627} and path optimization \cite{PhysRevResearch.2.023295,ding2021nonadiabatic,ding2022path,PhysRevApplied.19.044076,PhysRevApplied.21.064048,PhysRevApplied.22.014060}, etc. Notably, composite pulses improve the robustness through redundant trajectory correction, yet gate time is significantly prolonged. While the typical method that does not utilize any additional optimal control is path optimization, and it enhances the robustness only through reasonably shortening the evolution path. However, it either requires additional control of the detuning field \cite{ding2021nonadiabatic,PhysRevApplied.21.064048,PhysRevApplied.22.014060} or is limited to several certain specific quantum gates or/and for a certain type error source \cite{PhysRevResearch.2.023295,ding2022path,PhysRevApplied.19.044076} (not to universal robust for universal quantum gates with simple control). Therefore, it is necessary to find a solution that minimizes the gate times with simple pulse control while implementing universal robustness for any type of geometric gates.

In this work, we propose a NQGC scheme with shortened composite pulses, it can greatly improve the robustness of an arbitrary type of geometric gate against universal systematic errors by optimizing additional evolution parameters. More importantly, the gate times of our scheme are significantly reduced compared to conventional composite NGQC schemes, and even are smaller than conventional single-loop NGQC schemes for some specific gates. Numerical results verify that a universal geometric gate set (including two-qubit entangling gates) implemented by using our optimized method can show error-resilient features better than conventional single-loop NGQC gates. Finally, we propose an implementation of our proposal on superconducting transmon qubits using an experimentally demonstrated parametrically tunable coupling technique \cite{PhysRevA.96.062323,reagor2018demonstration,PhysRevApplied.10.034050,PhysRevApplied.10.054009}, where the gate performance advantage is still maintained. Therefore, our scheme provides a promising strategy for realizing high-fidelity robust geometric quantum gates, which is necessary to the large-scale quantum computation.

\section{Composite short-path NGQC}

\subsection{Nonadiabatic geometric phases}\label{sec2a}
\begin{figure}[tbp]
\centering
\includegraphics[width=1.0\linewidth]{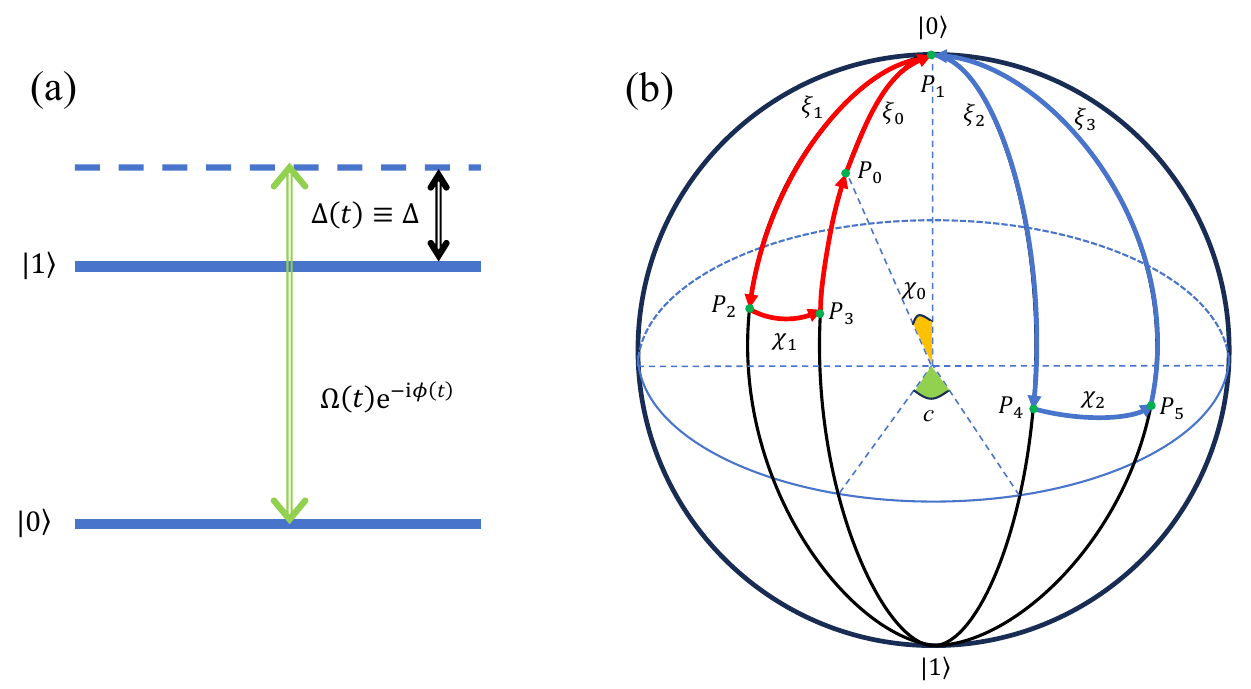}
\caption{Illustration of our scheme. (a) A simple qubit coupling structure driven by a microwave field with a constant detuning $\Delta$. (b) Illustration of a closed evolution trajectory of state $\Psi_1(t)$ for our single-qubit geometric gates in the Bloch sphere, which going along $P_0\rightarrow P_1\rightarrow P_2\rightarrow P_3\rightarrow P_0\rightarrow P_1\rightarrow P_4\rightarrow P_5\rightarrow P_1\rightarrow P_0$,  forming two triangle-cap-shaped structures with different sizes, where $\chi_1$, $\chi_2$ and $c$ are optimizable evolution parameters.}\label{Fig1}
\end{figure}

We begin by introducing a general framework of inducing the nonadiabatic geometric phases. In the interaction picture, we consider that a qubit is driven by a classical microwave field, whose interaction Hamiltonian can be expressed as (hereafter assuming $\hbar=1$)
\begin{equation}\label{eq1}
H_1(t)=\frac{1}{2}\left(
                  \begin{array}{cc}
                    -\Delta(t) & \Omega(t)e^{-i\phi(t)} \\
                    \Omega(t)e^{i\phi(t)} & \Delta(t) \\
                  \end{array}
                \right),
\end{equation}
in the computational subspace $\{|0\rangle,|1\rangle\}$, where $\Omega(t)$, $\phi(t)$ and $\Delta(t)$ are the time-dependent amplitude, phase and detuning of driving field, respectively, illustrated in Fig. \ref{Fig1}(a).

Here, to obtain the evolution operator, we choose the following a pair of orthogonal state vectors $\{\Psi_{1,2}(t)\}$:
\begin{eqnarray}
\begin{split}
|\Psi_1(t)\rangle&=e^{if_1(t)}\left[\cos\frac{\chi(t)}{2}|0\rangle+\sin\frac{\chi(t)}{2}e^{i\xi(t)}|1\rangle\right], \\
|\Psi_2(t)\rangle&=e^{if_2(t)}\left[\sin\frac{\chi(t)}{2}e^{-i\xi(t)}|0\rangle-\cos\frac{\chi(t)}{2}|1\rangle\right],
\end{split}
\end{eqnarray}
as the evolution states, whose evolution details on a Bloch sphere can be described by the time-dependent polar angle $\chi(t)$ and azimuth angle $\xi(t)$ as shown in Fig. \ref{Fig1}(b), and the accumulated total phases $f_k(t)$ satisfy $f_k(0)=0$ with $k=1,2$. Obviously, the dynamic process of evolution state $\Psi_k(t)$ is governed by the Hamiltonian $H_1(t)$ and connected by the Schr\"{o}dinger equation $i(\partial/\partial t)|\Psi_k(t)\rangle=H_1(t)|\Psi_k(t)\rangle$, and thus we can solve the correspondence linking the evolution parameters $\{\chi(t),\xi(t)\}$ to the control parameters $\{\Omega(t),\phi(t),\Delta(t)\}$ as
\begin{eqnarray}
\begin{split}
\dot{\chi}(t)&=\Omega(t)\sin[\phi(t)-\xi(t)],\\
\dot{\xi}(t)&=-\Delta(t)-\Omega(t)\cot{\chi(t)}\cos[\phi(t)-\xi(t)].
\end{split}
\end{eqnarray}
Note that the control parameters $\{\Omega(t),\phi(t),\Delta(t)\}$ of the Hamiltonian can be reconstructed via the above constraint relations, on the condition that given a prescribed evolution trajectory with known polar and azimuth angles. Therefore, after any cyclic evolution over duration $\tau$, the evolved states will be $U(\tau)|\Psi_k(0)\rangle=e^{if_k(\tau)}|\Psi_k(0)\rangle$, and thus the associated evolution operator can be calculated as
\begin{eqnarray}\label{eq4}
U(\tau)&=&\sum_{k=1,2}e^{if_k(\tau)}|\Psi_k(0)\rangle\langle \Psi_k(0)| \nonumber \\
&=&\left(
   \begin{array}{cc}
     \cos{\gamma}+i\sin{\gamma\cos\chi_0} & i\sin\gamma\sin\chi_0e^{-i\xi_0} \\
     i\sin\gamma\sin\chi_0e^{i\xi_0} & \cos{\gamma}-i\sin{\gamma\cos\chi_0}\\
   \end{array}
 \right), \nonumber\\
\end{eqnarray}
where $\chi_0=\chi(0)$ and $\xi_0=\xi(0)$ denote the initial polar and azimuth angles, respectively, and the $\gamma$ satisfies
\begin{equation}
\gamma=f_1(\tau)=-f_2(\tau)=\int_0^{\tau}\frac{\dot{\xi}(t)[1-\cos{\chi(t)}+\Delta(t)]}{2\cos{\chi(t)}}\mathrm{d}t,
\end{equation}
which corresponds to the accumulated total phase during the entire evolution process. Moreover, this phase can be divided into two parts, namely, the dynamical phase
\begin{eqnarray}
\gamma_d=-\int_0^{\tau}\langle \Psi_1(t)|H_1(t)|\Psi_1(t)\rangle \mathrm{d}t \nonumber \\
=\frac{1}{2}\int_0^{\tau}\frac{\dot{\xi}(t)\sin^2\chi(t)+\Delta(t)}{\cos\chi(t)}\mathrm{d}t,
\end{eqnarray}
and the remaining nonadiabatic geometric phase
\begin{equation}\label{eq7}
\gamma_g=\gamma-\gamma_d=-\frac{1}{2}\int_0^{\tau}\dot{\xi}(t)[1-\cos{\chi(t)}]\mathrm{d}t.
\end{equation}
Since it can be seen that from the Eq. (\ref{eq7}), geometric phase $\gamma_g$ is just equal to half of the solid angle enclosed by the evolution trajectory, the geometric nature \cite{PhysRevLett.58.1593} of which can be guaranteed. And then, geometric gates constructed by the geometric phase exhibit inherent fault-tolerant robustness against local errors. However, the existence of dynamical phase $\gamma_d$, associated with the specific evolution details, will undermine the global error resistance of the geometric gates as shown in Eq. (\ref{eq4}). Therefore, we here will aim to realize geometrically robust protection for the evolution process by eliminating the dynamical phase and maintain cyclic evolution simultaneously.
In addition, it is worth noting that for any given geometric gates, we only need to fix the boundary condition of $\chi(t)$ and $\xi(t)$, the specific shapes of $\chi(t)$ and $\xi(t)$ can be arbitrary as seen in Eq. (\ref{eq7}), which correspond to the arbitrary shapes closed evolution trajectories with different gate robustness. This provides more flexibility for optimizing the geometric gates.

\subsection{Gate construction for shortened composite geometric paths}
To eliminate the dynamical phase, here, we consider the case that utilizing the longitude and latitude lines as the basic configuration for simple pulse control. Specifically, on the longitude, since $\dot\xi(t)=0$, then setting the detuning will be zero, i.e., $\Delta(t)=0$ to lead to $\gamma_d=0$, while on the latitude line, $\dot\xi(t)\neq0$, and thus by setting the time-independent $\Delta(t)=\Delta=-(1/\tau)\int_0^{\tau}\dot{\xi}(t)\sin\chi(t)\mathrm{d}t$ with $\tau$ being the duration of evolution segment, to ensure that the dynamical phase is still zero. In any case, the setting of the constant detuning (including being zero) is an experiment-friendly control. Furthermore, the phase $\phi(t)$ is constant and is associated with the pulse shape on the longitude and latitude lines, respectively. The pulse shape $\Omega(t)$ can be arbitrary, and can be chosen as a simple form $\Omega_m\sin^2(\pi t/\tau)$ for experimental preference. As we all know, composite pulses, as a typical optimal control technology, can effectively enhance the robustness of geometric gates \cite{PhysRevApplied.10.054051,PhysRevA.103.032609}, but it will also significantly prolong the gate operation time. Moreover, owing to the fixed evolution paths, conventional composite pulse geometric schemes can only optimize specific types of noise, such as the Rabi error or the detuning error. Here, we propose to integrate path-optimized strategy and composite-pulse technique into the construction of geometric gates, aiming to maximize the suppression of gate time while maintaining strong universal robustness.

In detail, as shown in Fig. \ref{Fig1}(b), to achieve the optimized universal single-qubit geometric gates, the evolution path is divided into nine segments in the Bloch sphere, and the Hamiltonian parameters $\{\Omega(t),\phi(t),\Delta(t)\}$ for each segment can be set as follows:
\begin{subequations}\label{eq8}
\begin{eqnarray}
t\in[0,\tau_1]: \int_0^{\tau_1}\Omega(t)\mathrm{d}t=\chi_0,\Delta=0,\phi=\xi_0-\frac{\pi}{2},
\end{eqnarray}
\begin{eqnarray}
t\in[\tau_1,\tau_2]: \int_{\tau_1}^{\tau_2}\Omega(t)\mathrm{d}t=\chi_1,\Delta=0,\phi=\xi_1+\frac{\pi}{2},
\end{eqnarray}
\begin{eqnarray}
\begin{split}
&\quad t\in[\tau_2,\tau_3]: \int_{\tau_2}^{\tau_3}\Omega(t)\mathrm{d}t=|(\xi_1-\xi_0)\tan\chi_1\cos^2\chi_1|,   \\ &\;\quad\qquad\qquad\qquad\qquad\Delta=\frac{(\xi_1-\xi_0)\sin^2\chi_1}{\tau_3-\tau_2},  \\
&\quad\phi(t)= \left\{
            \begin{array}{l}
\xi_1+\frac{\int\Omega(t)\mathrm{d}t}{\tan{\chi_1}}-\Delta t+\pi,  \quad 0<\chi_1<\pi/2, \\
\xi_1-\frac{\int\Omega(t)\mathrm{d}t}{\tan{\chi_1}}-\Delta t,    \quad\pi/2<\chi_1\leq\pi,
            \end{array}
          \right.
\end{split}
\end{eqnarray}
\begin{eqnarray}
t\in[\tau_3,\tau_4]: \int_{\tau_3}^{\tau_4}\Omega(t)\mathrm{d}t=|\chi_1-\chi_0|,\quad\Delta=0,\nonumber \\
\phi=    \left\{ \begin{array}{l}
\xi_0-\frac{\pi}{2},  \quad \chi_1<\chi_0, \\
\xi_0+\frac{\pi}{2},  \quad \chi_1>\chi_0,
            \end{array}
          \right.
\end{eqnarray}
\begin{eqnarray}
t\in[\tau_4,\tau_5]: \int_{\tau_4}^{\tau_5}\Omega(t)\mathrm{d}t=\chi_0,\Delta=0,\phi=\xi_0-\frac{\pi}{2},
\end{eqnarray}
\begin{eqnarray}
t\in[\tau_5,\tau_6]: \int_{\tau_5}^{\tau_6}\Omega(t)\mathrm{d}t=\chi_2,\Delta=0,\phi=\xi_2+\frac{\pi}{2},
\end{eqnarray}
\begin{eqnarray}
\begin{split}
&\quad t\in[\tau_6,\tau_7]: \int_{\tau_6}^{\tau_7}\Omega(t)\mathrm{d}t=|(\xi_2-\xi_3)\tan\chi_2\cos^2\chi_2|,   \\ &\;\quad\qquad\qquad\qquad\qquad\Delta=\frac{(\xi_2-\xi_3)\sin^2\chi_2}{\tau_7-\tau_6},  \\
&\quad\phi(t)= \left\{
            \begin{array}{l}
\xi_2+\frac{\int\Omega(t)\mathrm{d}t}{\tan{\chi_2}}-\Delta t+\pi,  \quad 0<\chi_2<\pi/2, \\
\xi_2-\frac{\int\Omega(t)\mathrm{d}t}{\tan{\chi_2}}-\Delta t,    \quad\pi/2<\chi_2\leq\pi,
            \end{array}
          \right.
\end{split}
\end{eqnarray}
\begin{eqnarray}
t\in[\tau_7,\tau_8]: \int_{\tau_7}^{\tau_8}\Omega(t)\mathrm{d}t=\chi_2,\Delta=0,\phi=\xi_3-\frac{\pi}{2},
\end{eqnarray}
\begin{eqnarray}
t\in[\tau_8,\tau]: \int_{\tau_8}^{\tau}\Omega(t)\mathrm{d}t=\chi_0,\Delta=0,\phi=\xi_0+\frac{\pi}{2},
\end{eqnarray}
\end{subequations}
which corresponds to the evolution trajectory of the evolution state $|\Psi_1(t)\rangle$ carried out along the path $P_0\rightarrow P_1\rightarrow P_2\rightarrow P_3\rightarrow P_0\rightarrow P_1\rightarrow P_4\rightarrow P_5\rightarrow P_1\rightarrow P_0$, forming two triangle-cap-shaped structures with different sizes. Here, $\chi_1$ and $\chi_2$ represent the polar angle values on the third $(t\in[\tau_2, \tau_3])$ and seventh $(t\in[\tau_6, \tau_7])$ evolution segments, respectively, which correspond on the latitude line and thus is a constant. Therefore, at the final time $\tau$, the evolution operator can be calculated as
\begin{eqnarray}\label{eq9}
  U_1(\tau) &=& U_1(\tau,\tau_8)U_1(\tau_8,\tau_7)U_1(\tau_7,\tau_6)U_1(\tau_6,\tau_5)U_1(\tau_5,\tau_4) \nonumber \\
   &\cdot& U_1(\tau_4,\tau_3)U_1(\tau_3,\tau_2)U_1(\tau_2,\tau_1)U_1(\tau_1,0) \nonumber \\
   &=&\mathrm{e}^{\mathrm{i}\gamma_g\vec{n}\cdot\vec{\sigma}},
\end{eqnarray}
which represents an arbitrary single-qubit geometric gate, i.e., a geometric rotation around an arbitrary axis $\vec{n}$ by an arbitrary angle $-2\gamma_g$, with $\vec{n}=(\sin\chi_0\cos\xi_0,\sin\chi_0\sin\xi_0,\cos\chi_0)$ and the Pauli vector $\vec{\sigma}=(\sigma_x,\sigma_y,\sigma_z)$. Moreover, the geometric phase can be expressed as $\gamma_g=(\xi_1-\xi_0)(1-\cos\chi_1)=(\xi_0+c-\xi_3)(1-\cos\chi_2)$. It is clear that the implementation of arbitrary geometric single-qubit gates only requires fixing the values of $\chi_0$, $\xi_0$, and $\gamma_g$ from the Eq. (\ref{eq9}). Therefore, the free choice of the available evolution parameters $\chi_1$ and $\chi_2$ within the range of $[0, \pi]$ and $c$ within the range of $[0, 2\pi]$ allows for flexibility in the design of the control parameters $\{\Omega(t),\phi(t),\Delta(t)\}$, thereby providing us with more options to effectively correct the nine-segment geometric trajectory against systematic errors.

\begin{figure}[tbp]
\centering
\includegraphics[width=1.0\linewidth]{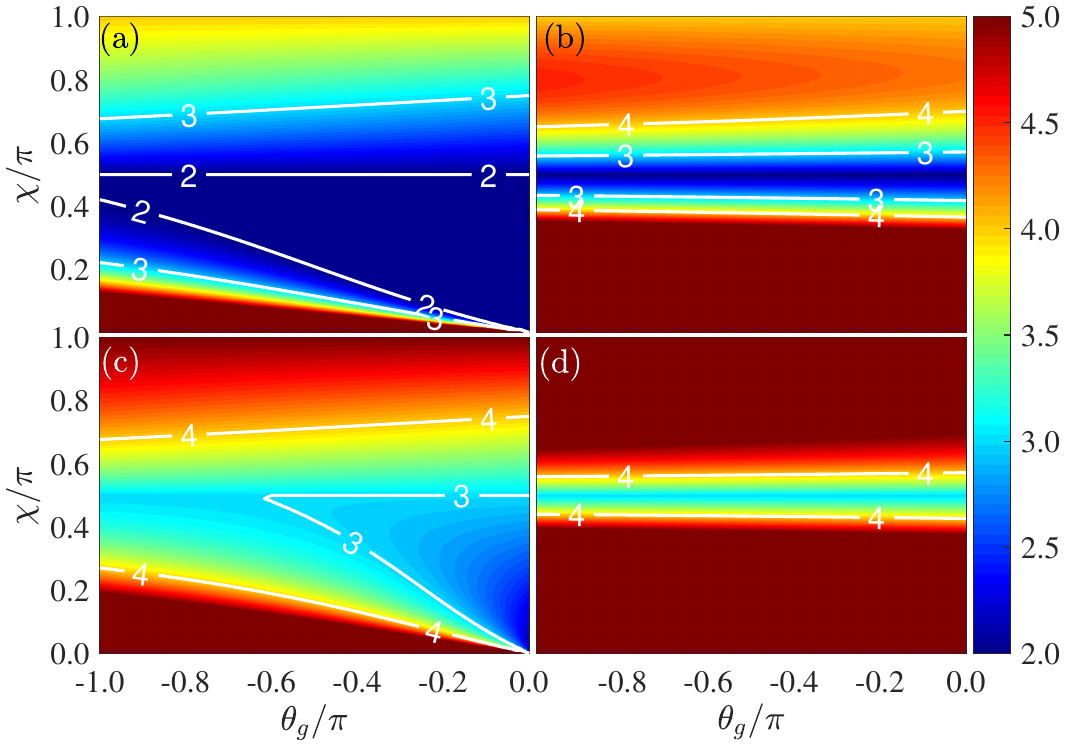}
\caption{The pulse area $\mathcal{S}=\int_0^{\tau}\Omega(t)\mathrm{d}t$ in units of $\pi$, as functions of the rotation angle $\theta_g$ and evolution parameter $\chi$ for the geometric rotation gates $Z^g_{-\theta_g}$ under $\gamma_g$ being (a) $\theta_g/2$ and (b) $\theta_g/2-2\pi$, respectively; that for geometric rotation gates $X^g_{-\theta_g}$ (or $Y^g_{-\theta_g}$) under $\gamma_g$ being (c) $\theta_g/2$ and (d) $\theta_g/2-2\pi$, respectively, where small pulse area means short evolution time.}\label{Fig2}
\end{figure}
Specifically, to implement the geometric rotation gates around $X$, $Y$ and $Z$ axes, denoted as $X_{-\theta_g}^g$, $Y_{-\theta_g}^g$ and $Z_{-\theta_g}^g$ with $\theta_g$ being negative angles, we can set $(\chi_0,\xi_0)=(\pi/2,0)$, $(\pi/2,\pi/2)$ and $(0,0)$, respectively, with the same $\gamma_g=\theta_g/2$ (or $\theta_g/2-2\pi$). In addition, we tested how the required gate time of these geometric gates varies with the evolution parameter $\chi$ and the rotation angle $\theta_g$, where the required gate time is defined as the total pulse area $\mathcal{S}=\int_0^{\tau}\Omega(t)\mathrm{d}t$ of all evolution segments, and $\chi=\chi_1=\chi_2$ for simplicity. As shown in Fig. \ref{Fig2}, it can be concluded that the pulse areas of geometric $Z_{-\theta_g}^g$ gates are less than that of geometric $X_{-\theta_g}^g$ and $Y_{-\theta_g}^g$ gates as a whole. Therefore, to pursue the goal of shorter gate time, we here choose a universal geometric gate set for $Z^g_{\pi/2}$, $Z^g_{\pi/4}$, denoted as $S$ and $T$, and Hadamard $H$ [by setting $(\chi_0,\xi_0)=(\pi/4,0)$ and $\gamma_g=-\pi/2$ (or $-5\pi/2$)] gates as typical example to optimize the robustness under different parameters $\chi$ and $c$, the details of which is shown later in Sec. \ref{sec2c}. In addition, it is worth noting that from the Figs. \ref{Fig2}(a) and \ref{Fig2}(b), in a fairly wide area, the pulse areas of these geometric gates are less than $4\pi$ that corresponds to those of the conventional composite nonadiabatic GQC schemes \cite{PhysRevApplied.10.054051,PhysRevA.103.032609,ding2025composite}, and thus these schemes are only a special case of our scheme.

\subsection{Gate robustness}\label{sec2c}
\begin{figure}[tbp]
\centering
\includegraphics[width=1.0\linewidth]{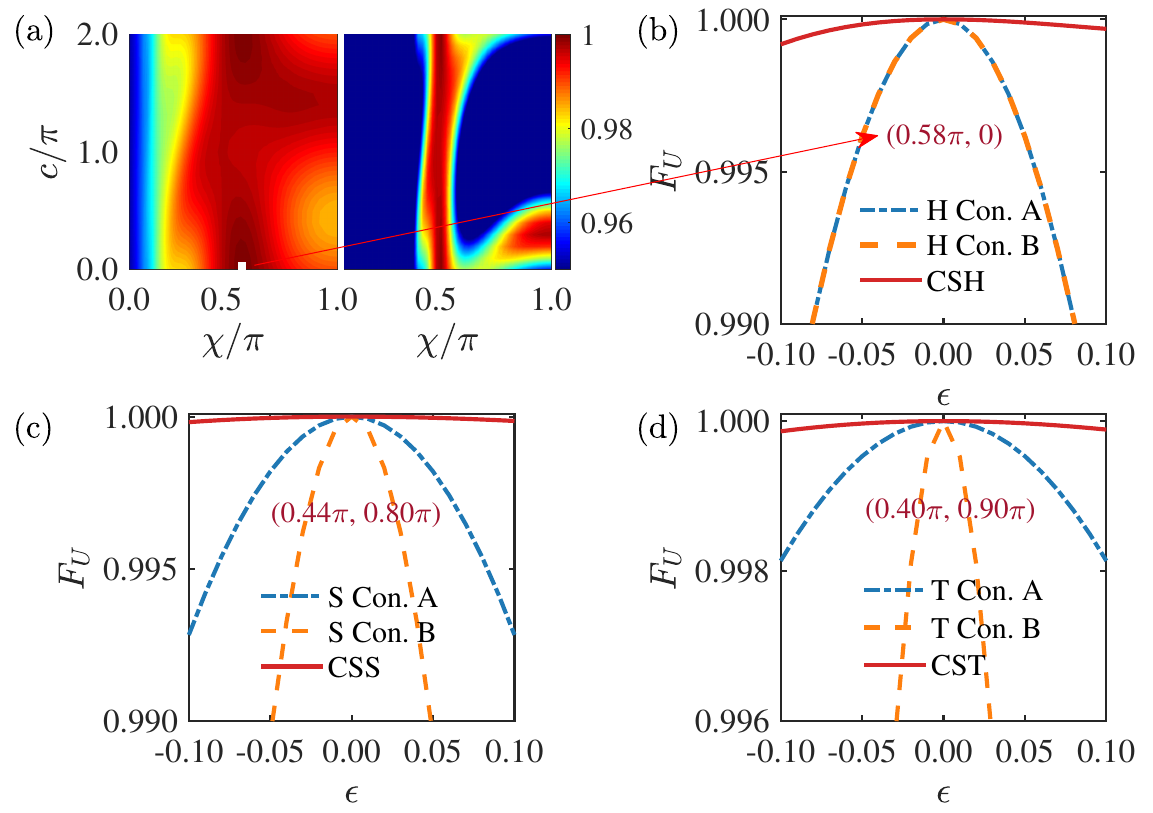}
\caption{As an example, (a) the gate fidelities of our CSH gate versus evolution parameters $\chi$ and $c$ to scan the optimal gate robustness against Rabi error with $\epsilon=0.1$, where the gate parameters $(\chi_0,\xi_0)=(\pi/4,0)$ and $\gamma_g=-\pi/2$ or $-5\pi/2$. (b) In the optimal regions of $\chi=0.58\pi$ and $c=0$, the robustness comparison between our CSH gate and the corresponding conventional single-loop geometric ones. In this same way, robust (c) CSS and (d) CST gates under respective optimal parameter regions have stronger error suppression than the corresponding conventional single-loop schemes.}\label{Fig3}
\end{figure}

Next, we will evaluate the robustness of the implemented geometric $H$, $S$ and $T$ gates against various systematic errors under the optimal evolution parameters. Note that the systematic errors typically include Rabi error, detuning error, and $Z\!Z$ crosstalk error. Here, we take Rabi error as a typical example to evaluate the robustness of these geometric gates. The assessments for other types of error sources can be found in Appendix A. Under the influence of the Rabi error, the pulse amplitude in Eq. (\ref{eq1}) will change as
\begin{equation}
\Omega(t) \rightarrow (1+\epsilon)\Omega(t),
\end{equation}
where $\epsilon$ denotes the error fraction in unit of the driving pulse $\Omega(t)$. Additionally, the gate fidelity with the error can be defined as \cite{PhysRevA.79.012105}
\begin{equation}
F^{\epsilon}_U=\frac{\mathrm{Tr}(U^{\dagger}U^{\epsilon})}{\mathrm{Tr}(U^{\dagger}U)},
\end{equation}
to evaluate the gate robustness, where $U$ and $U^{\epsilon}$ represent the ideal and error-affected evolution operators, respectively. In Fig. \ref{Fig3}(a), as a example, we show that the gate fidelities as a function of evolution parameters $\chi$ and $c$ with the error fraction $\epsilon=0.1$. We can clearly see that the gate fidelities are different under different $\chi$ and $c$, and thus we can find out the optimal parameter value $(\chi,c)=(0.58\pi,0)$ with the best gate robustness. Furthermore, we compare the gate robustness between our scheme and conventional single-loop geometric schemes, where the single-loop schemes have two different path configurations (denoted as Con.A and Con.B) with different gate robustness against Rabi error. Besides, the gate construction of single-loop geometric schemes is detailed in Appendix B. As shown in Fig. \ref{Fig3}(b), the gate robustness of our composite short-path geometric $H$ (CSH) gate against the Rabi error with $\epsilon\in[-0.1,0.1]$ is better than those of single-loop geometric case. Similarly, under the optimal evolution parameters $(\chi,c)=(0.44\pi,0.80\pi)$ and $(0.40\pi,0.90\pi)$, the superior error-resilient results for geometric $S$ and $T$ gates over the corresponding conventional single-loop geometric ones are shown in Figs. \ref{Fig3}(c) and \ref{Fig3}(d), respectively.

As an extension, in principle, we can release more available and free evolution parameters, i.e., not limiting $\chi_1=\chi_2$, to further improve the gate performance, and thus our scheme has the potential to effectively suppress a broader set of error sources simultaneously, or achieve superior suppression performance against a specific dominant errors. In addition, it is worth noting that on these optimal evolution parameters, the pulse areas of our CSH, CSS and CST gates can be calculated as $3.01\pi$, $1.87\pi$ and $1.71\pi$, which are all less than $4\pi$. Therefore, it can be concluded that our geometric scheme not only improves the gate robustness, but also significantly reduces the gate duration, compared with the conventional composite NGQC schemes.

\section{Physical Implementation}
In this section, we propose the physical implementation of our composite short-path geometric scheme on a 2D square lattice superconducting quantum circuit consisting of capacitively coupled transmon qubits \cite{PhysRevA.76.042319,PhysRevB.75.140515,krantz2019quantum}, as shown in Fig. \ref{Fig4}(a), and demonstrate its superior error-resilient characteristics compared to conventional single-loop geometric schemes.
\subsection{Universal single-qubit geometric gates}
\begin{figure}[tbp]
\centering
\includegraphics[width=1.0\linewidth]{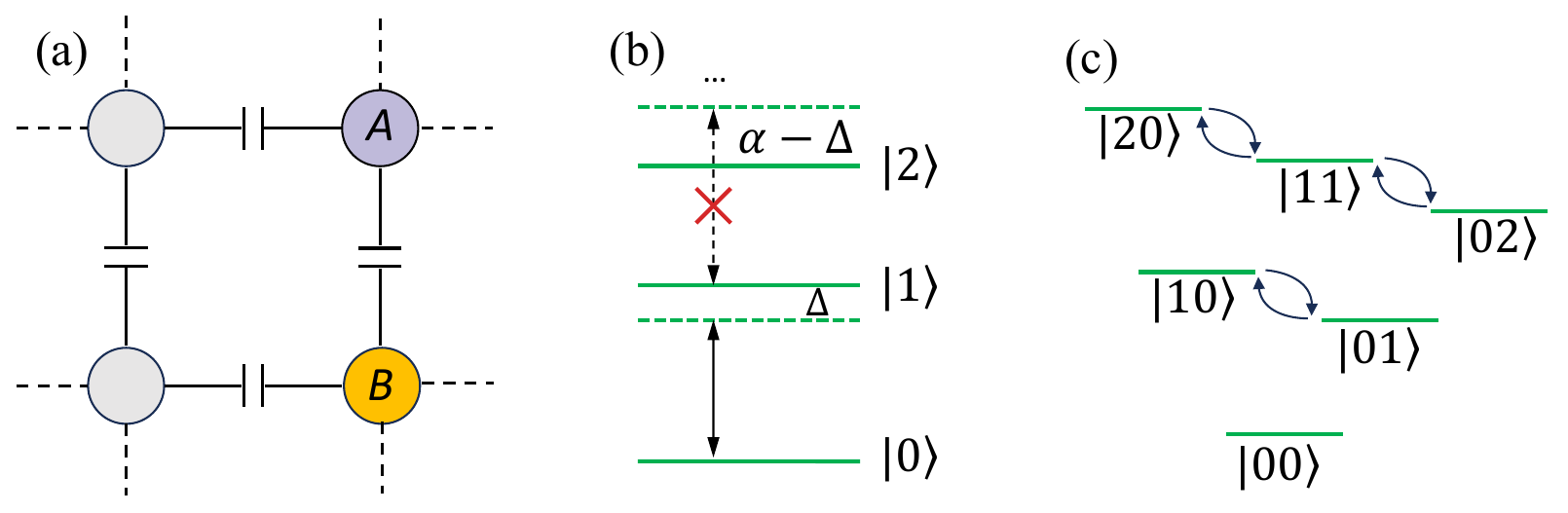}
\caption{Illustration of the proposed implementation. (a) Schematic diagram of a $2$D square transmon lattice connected by coupling
capacitively. Energy spectra for (b) a driven superconducting transmon with a weak anharmonicity $\alpha$ and a detuning $\Delta$, and (c) two
parametrically tunable coupled transmons $A$ and $B$, where different excitation subspaces can be addressed via parametrically tunable coupling.}\label{Fig4}
\end{figure}
Considering a transmon qubit is driven by an external microwave field, whose Hamiltonian can be described as
\begin{widetext}
\begin{equation}
\mathcal{H}_1(t)=\sum_{n=0}^{+\infty}\left(n\omega-\frac{n(n-1)}{2}\alpha\right)|n\rangle\langle n|+\frac{1}{2}\sum_{m=0}^{+\infty}\left\{\Omega(t)\exp\left[\mathrm{i}\left(\int_0^t\omega(t')\mathrm{d}t'-\phi(t)\right)\right]\sqrt{m}|m-1\rangle\langle m|+\textrm{H.c.}\right\},
\end{equation}
\end{widetext}
where $\alpha$ and $\omega$ are the anharmonicity and transition frequency of the transmon qubit, and $\Omega(t)$, $\omega(t)$ and $\phi(t)$ represent the time-dependent driving amplitude, frequency and phase of microwave field, respectively. Here, the lowest two levels $|0\rangle$ and $|1\rangle$ are regarded as our computational subspace, and due to the weak anharmonicity in transmons, the information will leaked out the computation subspace into the higher levels when controlling the qubit transition, which causes a decrease in gate fidelity. So, we consider these states as the leakage sources, in which level $|2\rangle$ as main leakage one. The corresponding energy spectrum is illustrated in Fig. \ref{Fig4}(b). Therefore, to improve the gate fidelity, we can correct the leakage error using a `derivative removal via adiabatic gate' technology \cite{PhysRevLett.103.110501,PhysRevLett.116.020501} with the corrected pulse shape
\begin{equation}
\Omega_c(t)=\Omega(t)-\mathrm{i}\frac{\dot{\Omega}(t)}{2\alpha}-\frac{\Delta(t)+\dot{\phi}(t)}{2\alpha},
\end{equation}
with $\Delta(t)=\omega-\omega(t)$, as the original pulse shape $\Omega(t)$ can be arbitrary one. And then, we can design a simple sinusoidal-like form $\Omega(t)=\Omega_m\sin^2(\pi t/\tau)$ for experiment-friendly. To apply the geometric evolution conditions in Eq. (\ref{eq8}), we transfer the Hamiltonian $\mathcal{H}_1(t)$ into a rotating framework with respect to frequency $\omega(t)$, and further map it into the computational subspace, the resulting Hamiltonian is equivalent to Eq. (\ref{eq1}). In addition, we utilize the Lindblad master equation \cite{lindblad1976generators} of
\begin{equation}
\dot{\rho}_1=\mathrm{i}\left[\rho_1,\mathcal{H}_1(t)\right]+\sum_{j=1}^{+\infty}\left(\frac{\kappa^1_-}{2}\mathcal{D}(\sigma_j)+\frac{\kappa^1_z}{2}\mathcal{D}(\chi_j)\right)
\end{equation}
to simulate numerically the gate fidelity under decoherence. Here, $\rho_1$ is the density matrix of transmon qubit, $\mathcal{D}(\mathcal{G})=2\mathcal{G}\rho_1\mathcal{G}^{\dagger}-\mathcal{G}^{\dagger}\mathcal{G}\rho_1-\rho_1\mathcal{G}^{\dagger}\mathcal{G}G$ is the Lindblad operator for operator $\mathcal{G}$, i.e., the standard lower operator $\sigma_j=\sqrt{j}|j-1\rangle\langle j|$ and the projector $\chi_j=j|j\rangle\langle j|$ for the $j$th energy level, and $\kappa_-^1$ and $\kappa_z^1$ are the relaxation and dephasing rates of the transmon, respectively. And, we use the formula $F_1^G=1/2\pi\int_0^{2\pi}\langle\psi_1^f|\rho_1|\psi_1^f\rangle\mathrm{d}\vartheta_1$ \cite{PhysRevLett.78.390} to assess the gate fidelity with the ideal final state of $|\psi_1^f\rangle=U|\psi_1^i\rangle$, the integration is numerically done for 1001 different initial states of $|\psi_1^i\rangle=\cos\vartheta_1|0\rangle+\sin\vartheta_1|1\rangle$ with $\vartheta_1$ being uniformly distributed within the range of $[0,2\pi]$.

\begin{figure*}[tbp]
\centering
\includegraphics[width=0.7\linewidth]{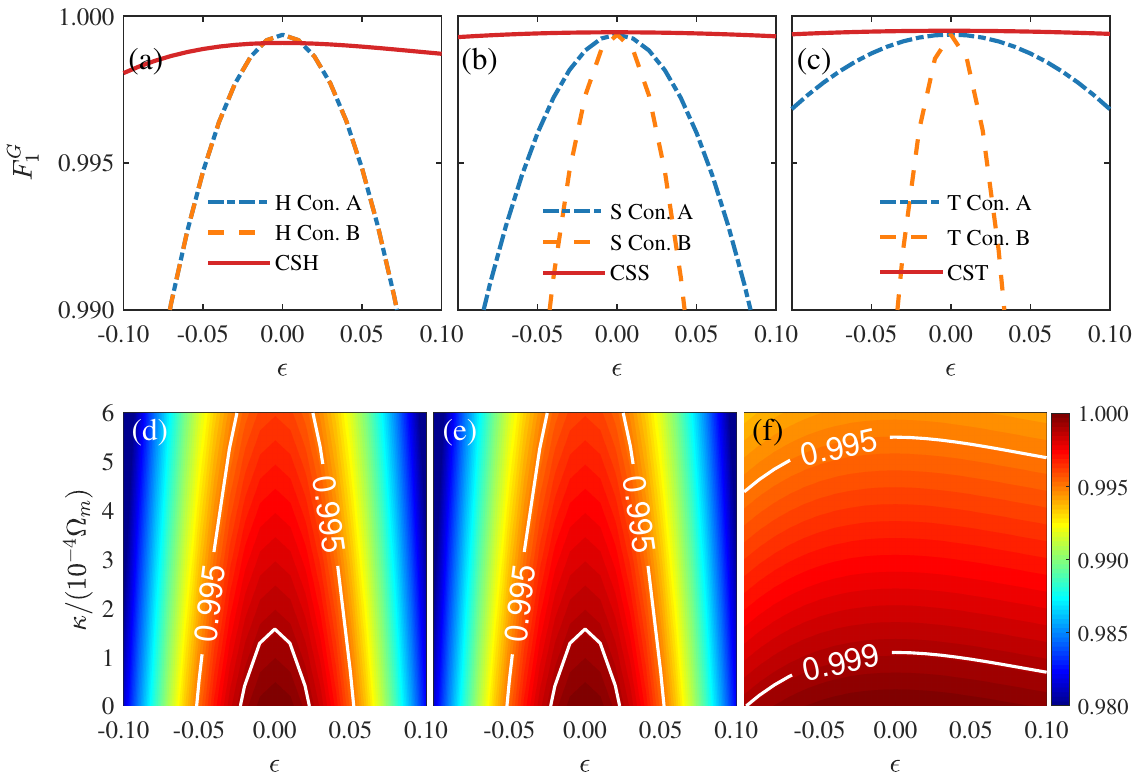}
\caption{(a)-(c) Comparison of robustness between our geometric scheme and conventional single-loop geometric schemes against Rabi error, under the decoherence rate $\kappa=\Omega_m/10^{4}$. Gate fidelities of $H$ gate as a function of the decoherence and Rabi error for conventional single-loop geometric schemes (d) in the configuration A and (e) in the configuration B, and (f) our geometric scheme.
}\label{Fig5}
\end{figure*}

In our simulations, to obtain the good performance for our geometric $\{H,S,T\}$ gates, we set $\alpha=2\pi\times320$ MHz, and the maximum of pulse $\Omega_m=2\pi\times\{35,78,70\}$ under the decoherence rates $\kappa^1_-=\kappa^1_z=\kappa=2\pi\times2$ kHz, which comes from state-of-the-art experiments \cite{kjaergaard2020superconducting,huang2020superconducting}. By these setting, the obtained gate fidelities can reach as $99.94\%$, $99.91\%$ and $99.95\%$, respectively. Through further numerical analysis, we find that the impacts of leakage error and decoherence on the gate fidelity are about $0.01\%$ and $0.05\%$, respectively, for geometric $H$ gate, and $0.08\%$ and $0.01\%$ for geometric $S$ gate, respectively, and $0.04\%$ and $0.01\%$ for geometric $T$ gate, respectively. Besides, the corresponding gate times are $86$ ns, $24$ ns and $24$ ns, respectively. In addition, with the decoherence rate $\kappa=\Omega_m/10^4$, the robustness comparisons between our scheme and single-loop ones are shown in Figs. \ref{Fig5}(a)-\ref{Fig5}(c). It can be seen that the robustness of our geometric gates still outperform largely corresponding single-loop schemes, except that the highest gate fidelity of the geometric $H$ gate is slightly decreased due to the long evolution time. To further reveal the overall performance of our geometric $H$ gate scheme, considering that under the different decoherence rates with $\kappa\in[0,6]\times10^{-4}\Omega_m$, the gate fidelities of $H$ gate as a function of the Rabi error $\epsilon$ as shown in Figs. \ref{Fig5}(d)-\ref{Fig5}(f). One can see that our geometric gate possesses a larger optimal area with a wider range of $\kappa$ and $\epsilon$.

\subsection{Nontrivial two-qubit geometric gates}
The design strategy of the above composite short-path can be readily applied to the construction for a nontrivial two-qubit geometric gate, which thus combine the universal single-qubit geometric gates to implement universal geometric quantum computation. Here, we consider a coupling configuration consisted of direct capacitive coupled transmon qubits, as shown in Fig. \ref{Fig4}(a). However, the coupling strengths between these transmons usually cannot be adjusted once their respective transition frequencies are fixed. To achieve controllable coupling of two transmon qubits, such as $A$ and $B$, we apply a well-controlled ac microwave on the qubit $B$ so that periodically modulating the transition frequency as $\omega_B(t)=\omega_B+\varepsilon\sin(\nu t+\varphi)$, where $\varepsilon$, $\nu$, and $\varphi$ are the microwave amplitude, frequency and phase, respectively. Then, the interaction Hamiltonian of these two transmon qubits can be written as
\begin{eqnarray}
\begin{aligned}
\mathcal{H}_{AB}(t)&=g_{AB}\Big\{|10\rangle\langle01|\mathrm{e}^{\mathrm{i}\Delta_1t}\mathrm{e}^{\mathrm{i}\beta\cos(\nu t+\varphi)}  \\
&\quad+\sqrt{2}|20\rangle\langle11|\mathrm{e}^{\mathrm{i}(\Delta_1-\alpha_A)t}\mathrm{e}^{\mathrm{i}\beta\cos(\nu t+\varphi)} \\
&\quad+\sqrt{2}|11\rangle\langle02|\mathrm{e}^{\mathrm{i}(\Delta_1+\alpha_B)t}\mathrm{e}^{\mathrm{i}\beta\cos(\nu t+\varphi)}+\textrm{H.c.}\Big\},
\end{aligned}
\end{eqnarray}
where $\beta=\varepsilon/\nu$, $\Delta_1=\omega_A-\omega_B$ and $g_{AB}$ are the frequency difference and the coupling strength between these two qubits, respectively. $\alpha_A$ and $\alpha_B$ are the corresponding intrinsic anharmonicities of the two qubits. We find this Hamiltonian can induce a different level transitions in the different subspaces, such as, $|10\rangle\leftrightarrow|01\rangle$ for the single-excitation subspace and $|20\rangle\leftrightarrow|11\rangle$ and $|11\rangle\leftrightarrow|02\rangle$ for the two-excitation subspace, as shown in Fig. \ref{Fig4}(c). Under the condition that driving the frequency $\nu$ meeting $\nu=\Delta_1$, the single-excitation subspace can be selected, and setting $\nu=\Delta_1+\alpha_B$ or $\nu=\Delta_1-\alpha_A$, we can pick up the two-excitation subspace. In this work we focus on utilizing subspace $\{|11\rangle,|02\rangle\}$ to implement a controlled phase gate owing to short evolution time.

\begin{figure}[tbp]
\centering
\includegraphics[width=1.0\linewidth]{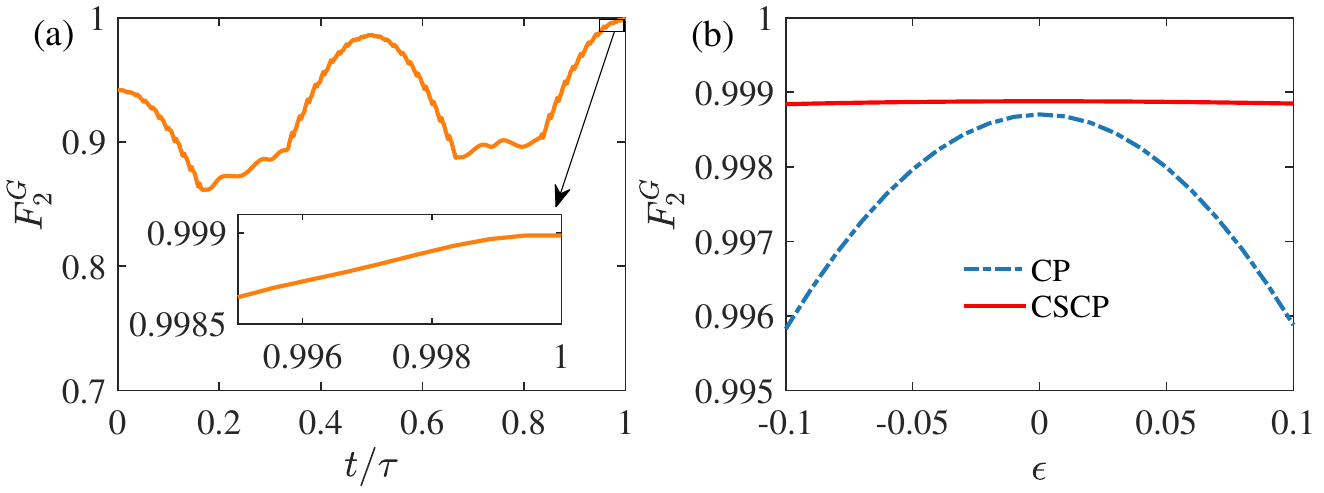}
\caption{(a) The evolution of gate fidelity of the implemented superconducting geometric controlled phase gate $(\gamma'_g=-\pi/4)$ versus time $t$. (b) Robustness comparison between our composite short-path geometric CP (CSCP) gate and conventional single-loop geometric CP gate against Rabi error, under the decoherence rate $\kappa=g'_{AB}/5000$.}\label{Fig6}
\end{figure}

Specifically, making a representation transformation with unitary operator $U(t)=\exp\left\{-\mathrm{i}\Delta't(|11\rangle\langle11|-|02\rangle\langle02|)/2\right\}$ on the $\mathcal{H}_{AB}(t)$ and utilizing the Jacobi-Anger identity of $\mathrm{e}^{\mathrm{i}\beta\cos(\nu t+\varphi)}=\sum_{-\infty}^{+\infty}\mathrm{i}^kJ_k(\beta)\exp[\mathrm{i}k(\nu t+\varphi)]$ with $J_k(\beta)$ being Bessel functions of the first kind. Thus, the Hamiltonian $\mathcal{H}_{AB}(t)$ will reduce to
\begin{equation}
\mathcal{H}_t(t)=\frac{1}{2}\left(
                              \begin{array}{cc}
                                -\Delta' & g'_{AB}\mathrm{e}^{-\mathrm{i}\varphi'} \\
                                g'_{AB}\mathrm{e}^{\mathrm{i}\varphi'} & \Delta' \\
                              \end{array}
                            \right),
\end{equation}
in a two-qubit subspace $\{|11\rangle,|02\rangle\}$, in which $\varphi'=\varphi-\pi/2$, $\Delta'=\nu-\Delta_1-\alpha_B$, $g'_{AB}=2\sqrt{2}g_{AB}J_1(\beta)$, and high-order oscillating terms have ignored by the rotating-wave approximation. Obviously, the above Hamiltonian $\mathcal{H}_t(t)$ has the same structure as that of Hamiltonian $H_1(t)$ in the Sec. \ref{sec2a}. Therefore, we can construct a two-qubit geometric controlled phase (CP) gate by implementing arbitrary single-qubit-like geometric phase gates in the subspace $\{|11\rangle,|02\rangle\}$ through appropriate parameter settings similar as that in Eq. (\ref{eq8}). As a result, the form of target geometric gates are
\begin{equation}
U_2(\tau)=\left(
            \begin{array}{cccc}
              1 & 0 & 0 & 0 \\
              0 & 1 & 0 & 0 \\
              0 & 0 & 1 & 0 \\
              0 & 0 & 0 & \mathrm{e}^{\mathrm{i}\gamma'_g} \\
            \end{array}
          \right),
\end{equation}
in the two-qubit computational subspace $\{|00\rangle,|01\rangle,|10\rangle,|11\rangle\}$. Alternatively, one can also obtain the controlled phase gates using the typical single-loop NGQC schemes \cite{PhysRevA.96.052316,PhysRevApplied.10.054051,PhysRevA.101.052302,PhysRevA.103.032609}. In the two-qubit scenario, the decoherence and the neglected high-frequency oscillation terms also need to be taken into account, as they can break the gate fidelity. To this end, we expand the master equation in a single-qubit case into a two-qubit one, as
\begin{equation}
\dot{\rho}_2=\mathrm{i}\left[\rho_2,\mathcal{H}_{AB}(t)\right]+\sum_{i=A,B}\sum_{j=1}^{+\infty}\left(\frac{\kappa^i_-}{2}\mathcal{D}(\sigma_j)+\frac{\kappa^i_z}{2}\mathcal{D}(\chi_j)\right)
\end{equation}
and using gate fidelity $F_2^G=1/4\pi^2\int_0^{2\pi}\int_0^{2\pi}\langle\psi_2^f|\rho_2|\psi_2^f\rangle\mathrm{d}\vartheta_1\mathrm{d}\vartheta_2$ to comprehensively evaluate gate performance, where $|\psi_2^f\rangle=U_2|\psi_2^i\rangle$ is the ideal final state with a general initial state of two coupled qubits being $|\psi_2^i\rangle=(\cos\theta_A|0\rangle_A+\sin\theta_A|1\rangle_A)\otimes(\cos\theta_B|0\rangle_B+\sin\theta_B|1\rangle_B)$. Here, we set $\gamma'_g=-\pi/4$ as a typical example, and the dynamics of gate fidelity as shown in Fig. \ref{Fig6}(a), where the optimal evolution parameters are $(\chi,c)=(0.44\pi,0.80\pi)$. One can see that the gate fidelity of our geometric CP gate can reach about $99.9\%$ under the experimentally accessible parameter settings \cite{kjaergaard2020superconducting} of $\beta=2.4$ and $\Delta_1=2\pi\times424$ MHz, and $\alpha_A=2\pi\times320$ MHz, $\alpha_B=2\pi\times300$ MHz, $g_{AB}=2\pi\times10$ MHz and $\kappa=\kappa_-^1=\kappa_z^1=\kappa_-^2=\kappa_z^2=2\pi\times2$ kHz. Through further analysis of the highest gate fidelity, we find that the impacts of higher-order oscillating terms and decoherence on gate fidelity are about $0.03\%$ and $0.07\%$ respectively, with corresponding gate time being $63.7$ ns. Besides, under the decoherence rate with $\kappa/g'_{AB}=2\times10^{-4}$, the comparison results of superior error-resilient two-qubit geometric gate against the Rabi error from the coupling strength deviation with $(1+\epsilon)g'_{AB}$ is shown in Fig. \ref{Fig6}(b).

\section{Conclusion}
In summary, we have developed a geometric scheme of universal robust control based on shortened composite pulses, with which the more robust geometric gates against systematic errors can be implemented by refining evolution trajectory, while significantly suppressing the accompanying evolution times compared to conventional composite schemes. Furthermore, our pulse control is simple and experiment friendly, as the pulse shapes are not set constraints in our construction of any type of geometric gates. And then, we propose an implementation of our scheme on superconducting transmon qubits, and demonstrate the robustness advantages over conventional NGQC schemes against universal errors. Therefore, our scheme provides an alternative method for implementing high-fidelity and robust GQC.

\begin{acknowledgements}
This work was supported by the Natural Scientific Research Foundation of Anhui Provincial Education Department (Grants No. 2025AHGXZK40448 and No. 2023AH050481), Anhui Province Higher Education Provincial Quality Engineering Project (Grant No. 2023zygzts036), Doctoral Research Startup Funding Project of Anqing Normal University (Grant No. 231055), and Provincial College Students' Innovation Training Program (Grant No. 120250084).

\end{acknowledgements}

\appendix
\section{Similar suppression effect for other error sources}
\begin{figure}[tbp]
\centering
\includegraphics[width=1.0\linewidth]{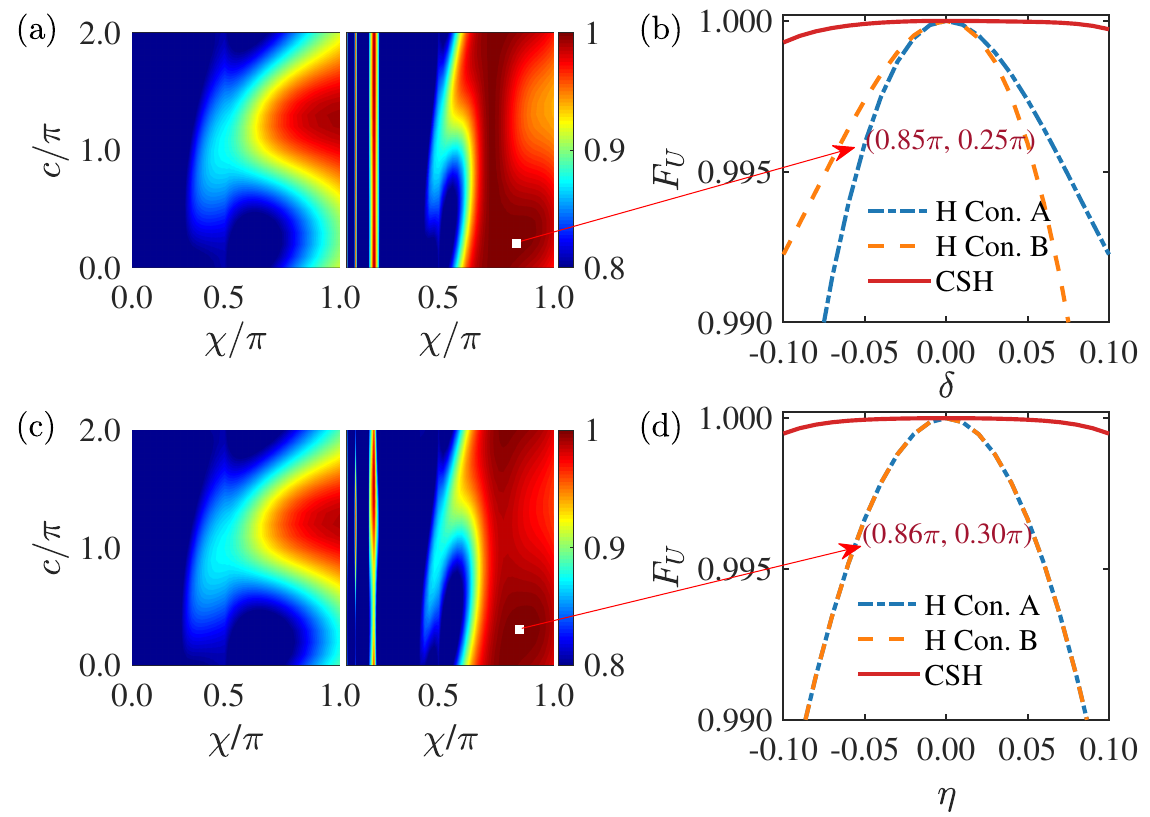}
\caption{The gate fidelity of our CSH gate to (a) detuning error and (c) $Z\!Z$ crosstalk under different evolution parameters $\chi$ and $c$, where the error quantities considered are $\delta=0.1$ and $\eta=0.1$, respectively. Our geometric $H$ gate can show better error-resilient features than conventional single-loop geometric $H$ gates in the configurations A and B, with the optimal regions (b) $(\chi,c)=(0.85\pi,0.25\pi)$ resisting detuning error and (d) $(0.86\pi,0.30\pi)$ resisting $Z\!Z$ crosstalk.}\label{Fig7}
\end{figure}

In the main text, we take the Rabi error as a typical example to verify our scheme have stronger robustness than conventional NGQC schemes by optimizing two or more evolution parameters. Actually, our method can also be used to suppress other types of error sources. In our superconducting implementation, besides the regarded Rabi error from pulse amplitude deviation, there are also two common sources of error, i.e., the detuning error from qubit-frequency drift induced by flux noise \cite{krantz2019quantum} and residual $Z\!Z$ crosstalk \cite{PhysRevLett.127.080505,PhysRevLett.129.040502}, whose error forms can be modeled as, respectively,
\begin{eqnarray}
  \Delta(t)&\rightarrow& \Delta(t)+\delta\Omega_m,\nonumber \\
  \mathcal{H}_{zz} &=& \frac{\eta}{2}\Omega_m\sigma_{z}^A\otimes\sigma_{z}^B,
\end{eqnarray}
with $\delta$ and $\eta$ being the corresponding error quantities. As shown in Fig. \ref{Fig7}, taking geometric $H$ gate as example, the results demonstrate that our scheme can also obviously suppress these two errors in the respective optimal evolution parameters $(\chi,c)=(0.85\pi,0.25\pi)$ and $(\chi,c)=(0.86\pi,0.30\pi)$ over corresponding conventional NGQC schemes.

\section{Conventional single-loop geometric schemes}
Here, we detail the construction of the nonadiabatic geometric gates by utilizing the conventional single-loop NQGC schemes \cite{PhysRevApplied.10.054051,PhysRevA.103.032609}. For a resonantly driven two-level quantum system, the interaction Hamiltonian in the interaction picture reads
\begin{equation}
H_c(t)=\left(
         \begin{array}{cc}
           0 & \Omega_c(t)\mathrm{e}^{-\mathrm{i}\phi_c} \\
           \Omega_c(t)\mathrm{e}^{\mathrm{i}\phi_c} & 0 \\
         \end{array}
       \right),
\end{equation}
in which $\Omega_c(t)=\Omega_m\sin^2(\pi t/T)$ and $\phi_c$ represent the amplitude and phase of the driving field, respectively. To induce a geometric phase, the entire evolution time $T$ is divided into three segments, the pulse area and phase $\phi_c$ of each segment need to satisfy the following condition:
\begin{equation}\label{B2}
\begin{cases}
\displaystyle \int_{0}^{T_1} \Omega_c(t)\mathrm{d}t = \theta_c, & \phi_c = \phi - \frac{\pi}{2}, \quad\qquad t\in[0,T_1], \\
\displaystyle \int_{T_1}^{T_2}\Omega_c(t)\mathrm{d}t = \pi, & \phi_c = \phi + \gamma + \frac{\pi}{2},\quad t\in[T_1,T_2], \\
\displaystyle \int_{T_2}^{T} \Omega_c(t) \mathrm{d}t = \pi - \theta_c, & \phi_c = \phi - \frac{\pi}{2},\quad\quad\quad t\in[T_2,T].
\end{cases}
\end{equation}
Consequently, at the finial time $T$, the corresponding evolution operator can be calculated as
\begin{eqnarray}\label{B3}
  U_c(T) &=& U_c(T,T_2)U_c(T_2,T_1)U_c(T_1,0) \nonumber \\
   &=& \cos\gamma+\mathrm{i}\sin\gamma\left(
                              \begin{array}{cc}
                                \cos\theta_c & \sin\theta_c\mathrm{e}^{-\mathrm{i}\phi} \\
                                \sin\theta_c\mathrm{e}^{\mathrm{i}\phi} & -\cos\theta_c \\
                              \end{array}
                            \right) \nonumber \\
    &=& \mathrm{e}^{\mathrm{i}\gamma\textbf{n}\cdot\mathbf{\sigma}},
\end{eqnarray}
where $\sigma=(\sigma_x,\sigma_y,\sigma_z)$ is a vector for the Pauli operators in the two-state system $\{|0\rangle,|1\rangle\}$, and $\textbf{n}=(\sin\theta_c\cos\phi,\sin\theta_c\sin\phi,\cos\theta_c)$ is a unit vector oriented along an arbitrary direction $\textbf{n}$ on the Bloch sphere. Clearly, the $U_c(T)$ denotes a rotation gate rotated about axis $\textbf{n}$ by an angle of $-2\gamma$. In addition, we note that the evolution operator can also be written as
\begin{equation}
U_c(T)=\mathrm{e}^{\mathrm{i}\gamma}|\mu_+\rangle\langle\mu_+|+\mathrm{e}^{-\mathrm{i}\gamma}|\mu_-\rangle\langle\mu_-|,
\end{equation}
where $|\mu_+\rangle=\cos(\theta_c/2)|0\rangle+\exp(\mathrm{i}\phi)\sin(\theta_c/2)|1\rangle$ and $|\mu_-\rangle=\exp(-\mathrm{i}\phi)\sin(\theta_c/2)|0\rangle-\cos(\theta_c/2)|1\rangle$ represent the eigenstates of $\textbf{n}\cdot\mathbf{\sigma}$. As the parallel-transport condition holds, namely $\langle\mu_{\pm}|U_c^{\dagger}H_c(t)U_c(t)|\mu_{\pm}\rangle=0$, $U_c(T)$ constitutes a geometric gate. Consequently, arbitrary single-qubit geometric gates can be implemented through suitable choices of the parameters $(\theta_c,\phi,\gamma)$. For example, the parameter sets $\{\pi/4,0,-\pi/2\}$, $\{0,0,-\pi/4\}$ and $\{0,0,-\pi/8\}$ yield the $H$ gate, $S$ gate and $T$ gate, respectively.

Remarkably, if we only change the phase $\phi_c$ of the second segment in Eq. (\ref{B2}) into $\phi-\gamma+\pi/2$, the resulting evolution operator will be $U_c(T)=-\mathrm{e}^{\mathrm{i}\gamma\textbf{n}\cdot\mathbf{\sigma}}$, which is equivalent to that in Eq. (\ref{B3}). However, the corresponding gate robustness for implementing the same geometric gate is different \cite{PhysRevLett.124.230503}. The reason is that they possess different evolution trajectories, and thus we term that the former is ``configuration A'', and the latter is ``configuration B''.


\begin{thebibliography}{53}%
\makeatletter
\providecommand \@ifxundefined [1]{%
 \@ifx{#1\undefined}
}%
\providecommand \@ifnum [1]{%
 \ifnum #1\expandafter \@firstoftwo
 \else \expandafter \@secondoftwo
 \fi
}%
\providecommand \@ifx [1]{%
 \ifx #1\expandafter \@firstoftwo
 \else \expandafter \@secondoftwo
 \fi
}%
\providecommand \natexlab [1]{#1}%
\providecommand \enquote  [1]{``#1''}%
\providecommand \bibnamefont  [1]{#1}%
\providecommand \bibfnamefont [1]{#1}%
\providecommand \citenamefont [1]{#1}%
\providecommand \href@noop [0]{\@secondoftwo}%
\providecommand \href [0]{\begingroup \@sanitize@url \@href}%
\providecommand \@href[1]{\@@startlink{#1}\@@href}%
\providecommand \@@href[1]{\endgroup#1\@@endlink}%
\providecommand \@sanitize@url [0]{\catcode `\\12\catcode `\$12\catcode
  `\&12\catcode `\#12\catcode `\^12\catcode `\_12\catcode `\%12\relax}%
\providecommand \@@startlink[1]{}%
\providecommand \@@endlink[0]{}%
\providecommand \url  [0]{\begingroup\@sanitize@url \@url }%
\providecommand \@url [1]{\endgroup\@href {#1}{\urlprefix }}%
\providecommand \urlprefix  [0]{URL }%
\providecommand \Eprint [0]{\href }%
\providecommand \doibase [0]{http://dx.doi.org/}%
\providecommand \selectlanguage [0]{\@gobble}%
\providecommand \bibinfo  [0]{\@secondoftwo}%
\providecommand \bibfield  [0]{\@secondoftwo}%
\providecommand \translation [1]{[#1]}%
\providecommand \BibitemOpen [0]{}%
\providecommand \bibitemStop [0]{}%
\providecommand \bibitemNoStop [0]{.\EOS\space}%
\providecommand \EOS [0]{\spacefactor3000\relax}%
\providecommand \BibitemShut  [1]{\csname bibitem#1\endcsname}%
\let\auto@bib@innerbib\@empty
\bibitem [{\citenamefont {Shor}(1997)}]{shor1997polynomial}%
  \BibitemOpen
  \bibfield  {author} {\bibinfo {author} {\bibfnamefont {P.~W.}\ \bibnamefont
  {Shor}},\ }\bibinfo {title} {Polynomial-time algorithms for prime
  factorization and discrete logarithms on a quantum computer},\ \href@noop {}
  {\bibfield  {journal} {\bibinfo  {journal} {SIAM J. Comput.}\ }\textbf
  {\bibinfo {volume} {26}},\ \bibinfo {pages} {1484} (\bibinfo {year}
  {1997})}\BibitemShut {NoStop}%
\bibitem [{\citenamefont {Grover}(1997)}]{PhysRevLett.79.325}%
  \BibitemOpen
  \bibfield  {author} {\bibinfo {author} {\bibfnamefont {L.~K.}\ \bibnamefont
  {Grover}},\ }\bibinfo {title} {Quantum Mechanics Helps in Searching for a
  Needle in a Haystack},\ \href {\doibase 10.1103/PhysRevLett.79.325}
  {\bibfield  {journal} {\bibinfo  {journal} {Phys. Rev. Lett.}\ }\textbf
  {\bibinfo {volume} {79}},\ \bibinfo {pages} {325} (\bibinfo {year}
  {1997})}\BibitemShut {NoStop}%
\bibitem [{\citenamefont {Lloyd}(1995)}]{PhysRevLett.75.346}%
  \BibitemOpen
  \bibfield  {author} {\bibinfo {author} {\bibfnamefont {S.}~\bibnamefont
  {Lloyd}},\ }\bibinfo {title} {Almost Any Quantum Logic Gate is Universal},\
  \href {\doibase 10.1103/PhysRevLett.75.346} {\bibfield  {journal} {\bibinfo
  {journal} {Phys. Rev. Lett.}\ }\textbf {\bibinfo {volume} {75}},\ \bibinfo
  {pages} {346} (\bibinfo {year} {1995})}\BibitemShut {NoStop}%
\bibitem [{\citenamefont {Barenco}\ \emph {et~al.}(1995)\citenamefont
  {Barenco}, \citenamefont {Bennett}, \citenamefont {Cleve}, \citenamefont
  {DiVincenzo}, \citenamefont {Margolus}, \citenamefont {Shor}, \citenamefont
  {Sleator}, \citenamefont {Smolin},\ and\ \citenamefont
  {Weinfurter}}]{PhysRevA.52.3457}%
  \BibitemOpen
  \bibfield  {author} {\bibinfo {author} {\bibfnamefont {A.}~\bibnamefont
  {Barenco}}, \bibinfo {author} {\bibfnamefont {C.~H.}\ \bibnamefont
  {Bennett}}, \bibinfo {author} {\bibfnamefont {R.}~\bibnamefont {Cleve}},
  \bibinfo {author} {\bibfnamefont {D.~P.}\ \bibnamefont {DiVincenzo}},
  \bibinfo {author} {\bibfnamefont {N.}~\bibnamefont {Margolus}}, \bibinfo
  {author} {\bibfnamefont {P.}~\bibnamefont {Shor}}, \bibinfo {author}
  {\bibfnamefont {T.}~\bibnamefont {Sleator}}, \bibinfo {author} {\bibfnamefont
  {J.~A.}\ \bibnamefont {Smolin}}, \ and\ \bibinfo {author} {\bibfnamefont
  {H.}~\bibnamefont {Weinfurter}},\ }\bibinfo {title} {Elementary gates for
  quantum computation},\ \href {\doibase 10.1103/PhysRevA.52.3457} {\bibfield
  {journal} {\bibinfo  {journal} {Phys. Rev. A}\ }\textbf {\bibinfo {volume}
  {52}},\ \bibinfo {pages} {3457} (\bibinfo {year} {1995})}\BibitemShut
  {NoStop}%
\bibitem [{\citenamefont {De~Chiara}\ and\ \citenamefont
  {Palma}(2003)}]{PhysRevLett.91.090404}%
  \BibitemOpen
  \bibfield  {author} {\bibinfo {author} {\bibfnamefont {G.}~\bibnamefont
  {De~Chiara}}\ and\ \bibinfo {author} {\bibfnamefont {G.~M.}\ \bibnamefont
  {Palma}},\ }\bibinfo {title} {Berry Phase for a Spin $1/2$ Particle in a
  Classical Fluctuating Field},\ \href {\doibase 10.1103/PhysRevLett.91.090404}
  {\bibfield  {journal} {\bibinfo  {journal} {Phys. Rev. Lett.}\ }\textbf
  {\bibinfo {volume} {91}},\ \bibinfo {pages} {090404} (\bibinfo {year}
  {2003})}\BibitemShut {NoStop}%
\bibitem [{\citenamefont {Zhu}\ and\ \citenamefont
  {Zanardi}(2005)}]{PhysRevA.72.020301}%
  \BibitemOpen
  \bibfield  {author} {\bibinfo {author} {\bibfnamefont {S.-L.}\ \bibnamefont
  {Zhu}}\ and\ \bibinfo {author} {\bibfnamefont {P.}~\bibnamefont {Zanardi}},\
  }\bibinfo {title} {Geometric quantum gates that are robust against stochastic
  control errors},\ \href {\doibase 10.1103/PhysRevA.72.020301} {\bibfield
  {journal} {\bibinfo  {journal} {Phys. Rev. A}\ }\textbf {\bibinfo {volume}
  {72}},\ \bibinfo {pages} {020301(R)} (\bibinfo {year} {2005})}\BibitemShut
  {NoStop}%
\bibitem [{\citenamefont {Filipp}\ \emph {et~al.}(2009)\citenamefont {Filipp},
  \citenamefont {Klepp}, \citenamefont {Hasegawa}, \citenamefont
  {Plonka-Spehr}, \citenamefont {Schmidt}, \citenamefont {Geltenbort},\ and\
  \citenamefont {Rauch}}]{PhysRevLett.102.030404}%
  \BibitemOpen
  \bibfield  {author} {\bibinfo {author} {\bibfnamefont {S.}~\bibnamefont
  {Filipp}}, \bibinfo {author} {\bibfnamefont {J.}~\bibnamefont {Klepp}},
  \bibinfo {author} {\bibfnamefont {Y.}~\bibnamefont {Hasegawa}}, \bibinfo
  {author} {\bibfnamefont {C.}~\bibnamefont {Plonka-Spehr}}, \bibinfo {author}
  {\bibfnamefont {U.}~\bibnamefont {Schmidt}}, \bibinfo {author} {\bibfnamefont
  {P.}~\bibnamefont {Geltenbort}}, \ and\ \bibinfo {author} {\bibfnamefont
  {H.}~\bibnamefont {Rauch}},\ }\bibinfo {title} {Experimental Demonstration of
  the Stability of Berry's Phase for a Spin-$1/2$ Particle},\ \href {\doibase
  10.1103/PhysRevLett.102.030404} {\bibfield  {journal} {\bibinfo  {journal}
  {Phys. Rev. Lett.}\ }\textbf {\bibinfo {volume} {102}},\ \bibinfo {pages}
  {030404} (\bibinfo {year} {2009})}\BibitemShut {NoStop}%
\bibitem [{\citenamefont {Berry}(1984)}]{berry1984quantal}%
  \BibitemOpen
  \bibfield  {author} {\bibinfo {author} {\bibfnamefont {M.~V.}\ \bibnamefont
  {Berry}},\ }\bibinfo {title} {Quantal phase factors accompanying adiabatic
  changes},\ \href@noop {} {\bibfield  {journal} {\bibinfo  {journal} {Proc. R.
  Soc. A}\ }\textbf {\bibinfo {volume} {392}},\ \bibinfo {pages} {45} (\bibinfo
  {year} {1984})}\BibitemShut {NoStop}%
\bibitem [{\citenamefont {Jones}\ \emph {et~al.}(2000)\citenamefont {Jones},
  \citenamefont {Vedral}, \citenamefont {Ekert},\ and\ \citenamefont
  {Castagnoli}}]{jones2000geometric}%
  \BibitemOpen
  \bibfield  {author} {\bibinfo {author} {\bibfnamefont {J.~A.}\ \bibnamefont
  {Jones}}, \bibinfo {author} {\bibfnamefont {V.}~\bibnamefont {Vedral}},
  \bibinfo {author} {\bibfnamefont {A.}~\bibnamefont {Ekert}}, \ and\ \bibinfo
  {author} {\bibfnamefont {G.}~\bibnamefont {Castagnoli}},\ }\bibinfo {title}
  {Geometric quantum computation using nuclear magnetic resonance},\ \href@noop
  {} {\bibfield  {journal} {\bibinfo  {journal} {Nature}\ }\textbf {\bibinfo
  {volume} {403}},\ \bibinfo {pages} {869} (\bibinfo {year}
  {2000})}\BibitemShut {NoStop}%
\bibitem [{\citenamefont {Wu}\ \emph {et~al.}(2013)\citenamefont {Wu},
  \citenamefont {Gauger}, \citenamefont {George}, \citenamefont
  {M{\"o}tt{\"o}nen}, \citenamefont {Riemann}, \citenamefont {Abrosimov},
  \citenamefont {Becker}, \citenamefont {Pohl}, \citenamefont {Itoh},
  \citenamefont {Thewalt},\ and\ \citenamefont {Morton}}]{PhysRevA.87.032326}%
  \BibitemOpen
  \bibfield  {author} {\bibinfo {author} {\bibfnamefont {H.}~\bibnamefont
  {Wu}}, \bibinfo {author} {\bibfnamefont {E.~M.}\ \bibnamefont {Gauger}},
  \bibinfo {author} {\bibfnamefont {R.~E.}\ \bibnamefont {George}}, \bibinfo
  {author} {\bibfnamefont {M.}~\bibnamefont {M{\"o}tt{\"o}nen}}, \bibinfo
  {author} {\bibfnamefont {H.}~\bibnamefont {Riemann}}, \bibinfo {author}
  {\bibfnamefont {N.~V.}\ \bibnamefont {Abrosimov}}, \bibinfo {author}
  {\bibfnamefont {P.}~\bibnamefont {Becker}}, \bibinfo {author} {\bibfnamefont
  {H.-J.}\ \bibnamefont {Pohl}}, \bibinfo {author} {\bibfnamefont {K.~M.}\
  \bibnamefont {Itoh}}, \bibinfo {author} {\bibfnamefont {M.~L.~W.}\
  \bibnamefont {Thewalt}}, \ and\ \bibinfo {author} {\bibfnamefont {J.~J.~L.}\
  \bibnamefont {Morton}},\ }\bibinfo {title} {Geometric phase gates with
  adiabatic control in electron spin resonance},\ \href {\doibase
  10.1103/PhysRevA.87.032326} {\bibfield  {journal} {\bibinfo  {journal} {Phys.
  Rev. A}\ }\textbf {\bibinfo {volume} {87}},\ \bibinfo {pages} {032326}
  (\bibinfo {year} {2013})}\BibitemShut {NoStop}%
\bibitem [{\citenamefont {Huang}\ \emph {et~al.}(2019)\citenamefont {Huang},
  \citenamefont {Wu}, \citenamefont {Wang}, \citenamefont {Hou}, \citenamefont
  {Wang}, \citenamefont {Zhang}, \citenamefont {Lian}, \citenamefont {Liu},
  \citenamefont {Wang}, \citenamefont {Zhang}, \citenamefont {He},
  \citenamefont {Chang}, \citenamefont {Xu},\ and\ \citenamefont
  {Duan}}]{PhysRevLett.122.010503}%
  \BibitemOpen
  \bibfield  {author} {\bibinfo {author} {\bibfnamefont {Y.-Y.}\ \bibnamefont
  {Huang}}, \bibinfo {author} {\bibfnamefont {Y.-K.}\ \bibnamefont {Wu}},
  \bibinfo {author} {\bibfnamefont {F.}~\bibnamefont {Wang}}, \bibinfo {author}
  {\bibfnamefont {P.-Y.}\ \bibnamefont {Hou}}, \bibinfo {author} {\bibfnamefont
  {W.-B.}\ \bibnamefont {Wang}}, \bibinfo {author} {\bibfnamefont {W.-G.}\
  \bibnamefont {Zhang}}, \bibinfo {author} {\bibfnamefont {W.-Q.}\ \bibnamefont
  {Lian}}, \bibinfo {author} {\bibfnamefont {Y.-Q.}\ \bibnamefont {Liu}},
  \bibinfo {author} {\bibfnamefont {H.-Y.}\ \bibnamefont {Wang}}, \bibinfo
  {author} {\bibfnamefont {H.-Y.}\ \bibnamefont {Zhang}}, \bibinfo {author}
  {\bibfnamefont {L.}~\bibnamefont {He}}, \bibinfo {author} {\bibfnamefont
  {X.-Y.}\ \bibnamefont {Chang}}, \bibinfo {author} {\bibfnamefont
  {Y.}~\bibnamefont {Xu}}, \ and\ \bibinfo {author} {\bibfnamefont {L.-M.}\
  \bibnamefont {Duan}},\ }\bibinfo {title} {Experimental Realization of Robust
  Geometric Quantum Gates with Solid-State Spins},\ \href {\doibase
  10.1103/PhysRevLett.122.010503} {\bibfield  {journal} {\bibinfo  {journal}
  {Phys. Rev. Lett.}\ }\textbf {\bibinfo {volume} {122}},\ \bibinfo {pages}
  {010503} (\bibinfo {year} {2019})}\BibitemShut {NoStop}%
\bibitem [{\citenamefont {Aharonov}\ and\ \citenamefont
  {Anandan}(1987)}]{PhysRevLett.58.1593}%
  \BibitemOpen
  \bibfield  {author} {\bibinfo {author} {\bibfnamefont {Y.}~\bibnamefont
  {Aharonov}}\ and\ \bibinfo {author} {\bibfnamefont {J.}~\bibnamefont
  {Anandan}},\ }\bibinfo {title} {Phase change during a cyclic quantum
  evolution},\ \href {\doibase 10.1103/PhysRevLett.58.1593} {\bibfield
  {journal} {\bibinfo  {journal} {Phys. Rev. Lett.}\ }\textbf {\bibinfo
  {volume} {58}},\ \bibinfo {pages} {1593} (\bibinfo {year}
  {1987})}\BibitemShut {NoStop}%
\bibitem [{\citenamefont {Xiang-Bin}\ and\ \citenamefont
  {Keiji}(2001)}]{PhysRevLett.87.097901}%
  \BibitemOpen
  \bibfield  {author} {\bibinfo {author} {\bibfnamefont {W.}~\bibnamefont
  {Xiang-Bin}}\ and\ \bibinfo {author} {\bibfnamefont {M.}~\bibnamefont
  {Keiji}},\ }\bibinfo {title} {Nonadiabatic Conditional Geometric Phase Shift
  with NMR},\ \href {\doibase 10.1103/PhysRevLett.87.097901} {\bibfield
  {journal} {\bibinfo  {journal} {Phys. Rev. Lett.}\ }\textbf {\bibinfo
  {volume} {87}},\ \bibinfo {pages} {097901} (\bibinfo {year}
  {2001})}\BibitemShut {NoStop}%
\bibitem [{\citenamefont {Zhu}\ and\ \citenamefont
  {Wang}(2002)}]{PhysRevLett.89.097902}%
  \BibitemOpen
  \bibfield  {author} {\bibinfo {author} {\bibfnamefont {S.-L.}\ \bibnamefont
  {Zhu}}\ and\ \bibinfo {author} {\bibfnamefont {Z.~D.}\ \bibnamefont {Wang}},\
  }\bibinfo {title} {Implementation of Universal Quantum Gates Based on
  Nonadiabatic Geometric Phases},\ \href {\doibase
  10.1103/PhysRevLett.89.097902} {\bibfield  {journal} {\bibinfo  {journal}
  {Phys. Rev. Lett.}\ }\textbf {\bibinfo {volume} {89}},\ \bibinfo {pages}
  {097902} (\bibinfo {year} {2002})}\BibitemShut {NoStop}%
\bibitem [{\citenamefont {Zhu}\ and\ \citenamefont
  {Wang}(2003)}]{PhysRevLett.91.187902}%
  \BibitemOpen
  \bibfield  {author} {\bibinfo {author} {\bibfnamefont {S.-L.}\ \bibnamefont
  {Zhu}}\ and\ \bibinfo {author} {\bibfnamefont {Z.~D.}\ \bibnamefont {Wang}},\
  }\bibinfo {title} {Unconventional Geometric Quantum Computation},\ \href
  {\doibase 10.1103/PhysRevLett.91.187902} {\bibfield  {journal} {\bibinfo
  {journal} {Phys. Rev. Lett.}\ }\textbf {\bibinfo {volume} {91}},\ \bibinfo
  {pages} {187902} (\bibinfo {year} {2003})}\BibitemShut {NoStop}%
\bibitem [{\citenamefont {Leibfried}\ \emph {et~al.}(2003)\citenamefont
  {Leibfried}, \citenamefont {DeMarco}, \citenamefont {Meyer}, \citenamefont
  {Lucas}, \citenamefont {Barrett}, \citenamefont {Britton}, \citenamefont
  {Itano}, \citenamefont {Jelenkovi{\'c}}, \citenamefont {Langer},
  \citenamefont {Rosenband} \emph {et~al.}}]{leibfried2003experimental}%
  \BibitemOpen
  \bibfield  {author} {\bibinfo {author} {\bibfnamefont {D.}~\bibnamefont
  {Leibfried}}, \bibinfo {author} {\bibfnamefont {B.}~\bibnamefont {DeMarco}},
  \bibinfo {author} {\bibfnamefont {V.}~\bibnamefont {Meyer}}, \bibinfo
  {author} {\bibfnamefont {D.}~\bibnamefont {Lucas}}, \bibinfo {author}
  {\bibfnamefont {M.}~\bibnamefont {Barrett}}, \bibinfo {author} {\bibfnamefont
  {J.}~\bibnamefont {Britton}}, \bibinfo {author} {\bibfnamefont {W.~M.}\
  \bibnamefont {Itano}}, \bibinfo {author} {\bibfnamefont {B.}~\bibnamefont
  {Jelenkovi{\'c}}}, \bibinfo {author} {\bibfnamefont {C.}~\bibnamefont
  {Langer}}, \bibinfo {author} {\bibfnamefont {T.}~\bibnamefont {Rosenband}},
  \emph {et~al.},\ }\bibinfo {title} {Experimental demonstration of a robust,
  high-fidelity geometric two ion-qubit phase gate},\ \href@noop {} {\bibfield
  {journal} {\bibinfo  {journal} {Nature}\ }\textbf {\bibinfo {volume} {422}},\
  \bibinfo {pages} {412} (\bibinfo {year} {2003})}\BibitemShut {NoStop}%
\bibitem [{\citenamefont {Du}\ \emph {et~al.}(2006)\citenamefont {Du},
  \citenamefont {Zou},\ and\ \citenamefont {Wang}}]{PhysRevA.74.020302}%
  \BibitemOpen
  \bibfield  {author} {\bibinfo {author} {\bibfnamefont {J.}~\bibnamefont
  {Du}}, \bibinfo {author} {\bibfnamefont {P.}~\bibnamefont {Zou}}, \ and\
  \bibinfo {author} {\bibfnamefont {Z.~D.}\ \bibnamefont {Wang}},\ }\bibinfo
  {title} {Experimental implementation of high-fidelity unconventional
  geometric quantum gates using an NMR interferometer},\ \href {\doibase
  10.1103/PhysRevA.74.020302} {\bibfield  {journal} {\bibinfo  {journal} {Phys.
  Rev. A}\ }\textbf {\bibinfo {volume} {74}},\ \bibinfo {pages} {020302(R)}
  (\bibinfo {year} {2006})}\BibitemShut {NoStop}%
\bibitem [{\citenamefont {Song}\ \emph {et~al.}(2017)\citenamefont {Song},
  \citenamefont {Zheng}, \citenamefont {Zhang}, \citenamefont {Xu},
  \citenamefont {Zhang}, \citenamefont {Guo}, \citenamefont {Liu},
  \citenamefont {Xu}, \citenamefont {Deng}, \citenamefont {Huang} \emph
  {et~al.}}]{song2017continuous}%
  \BibitemOpen
  \bibfield  {author} {\bibinfo {author} {\bibfnamefont {C.}~\bibnamefont
  {Song}}, \bibinfo {author} {\bibfnamefont {S.-B.}\ \bibnamefont {Zheng}},
  \bibinfo {author} {\bibfnamefont {P.}~\bibnamefont {Zhang}}, \bibinfo
  {author} {\bibfnamefont {K.}~\bibnamefont {Xu}}, \bibinfo {author}
  {\bibfnamefont {L.}~\bibnamefont {Zhang}}, \bibinfo {author} {\bibfnamefont
  {Q.}~\bibnamefont {Guo}}, \bibinfo {author} {\bibfnamefont {W.}~\bibnamefont
  {Liu}}, \bibinfo {author} {\bibfnamefont {D.}~\bibnamefont {Xu}}, \bibinfo
  {author} {\bibfnamefont {H.}~\bibnamefont {Deng}}, \bibinfo {author}
  {\bibfnamefont {K.}~\bibnamefont {Huang}},  \emph {et~al.},\ }\bibinfo
  {title} {Continuous-variable geometric phase and its manipulation for quantum
  computation in a superconducting circuit},\ \href@noop {} {\bibfield
  {journal} {\bibinfo  {journal} {Nat. commun.}\ }\textbf {\bibinfo {volume}
  {8}},\ \bibinfo {pages} {1061} (\bibinfo {year} {2017})}\BibitemShut
  {NoStop}%
\bibitem [{\citenamefont {Zhao}\ \emph {et~al.}(2017)\citenamefont {Zhao},
  \citenamefont {Cui}, \citenamefont {Xu}, \citenamefont {Sj{\"o}qvist},\ and\
  \citenamefont {Tong}}]{PhysRevA.96.052316}%
  \BibitemOpen
  \bibfield  {author} {\bibinfo {author} {\bibfnamefont {P.~Z.}\ \bibnamefont
  {Zhao}}, \bibinfo {author} {\bibfnamefont {X.-D.}\ \bibnamefont {Cui}},
  \bibinfo {author} {\bibfnamefont {G.~F.}\ \bibnamefont {Xu}}, \bibinfo
  {author} {\bibfnamefont {E.}~\bibnamefont {Sj{\"o}qvist}}, \ and\ \bibinfo
  {author} {\bibfnamefont {D.~M.}\ \bibnamefont {Tong}},\ }\bibinfo {title}
  {Rydberg-atom-based scheme of nonadiabatic geometric quantum computation},\
  \href {\doibase 10.1103/PhysRevA.96.052316} {\bibfield  {journal} {\bibinfo
  {journal} {Phys. Rev. A}\ }\textbf {\bibinfo {volume} {96}},\ \bibinfo
  {pages} {052316} (\bibinfo {year} {2017})}\BibitemShut {NoStop}%
\bibitem [{\citenamefont {Chen}\ and\ \citenamefont
  {Xue}(2018)}]{PhysRevApplied.10.054051}%
  \BibitemOpen
  \bibfield  {author} {\bibinfo {author} {\bibfnamefont {T.}~\bibnamefont
  {Chen}}\ and\ \bibinfo {author} {\bibfnamefont {Z.-Y.}\ \bibnamefont {Xue}},\
  }\bibinfo {title} {Nonadiabatic Geometric Quantum Computation with
  Parametrically Tunable Coupling},\ \href {\doibase
  10.1103/PhysRevApplied.10.054051} {\bibfield  {journal} {\bibinfo  {journal}
  {Phys. Rev. Appl.}\ }\textbf {\bibinfo {volume} {10}},\ \bibinfo {pages}
  {054051} (\bibinfo {year} {2018})}\BibitemShut {NoStop}%
\bibitem [{\citenamefont {Zhang}\ \emph {et~al.}(2020)\citenamefont {Zhang},
  \citenamefont {Chen}, \citenamefont {Li}, \citenamefont {Wang},\ and\
  \citenamefont {Xue}}]{PhysRevA.101.052302}%
  \BibitemOpen
  \bibfield  {author} {\bibinfo {author} {\bibfnamefont {C.}~\bibnamefont
  {Zhang}}, \bibinfo {author} {\bibfnamefont {T.}~\bibnamefont {Chen}},
  \bibinfo {author} {\bibfnamefont {S.}~\bibnamefont {Li}}, \bibinfo {author}
  {\bibfnamefont {X.}~\bibnamefont {Wang}}, \ and\ \bibinfo {author}
  {\bibfnamefont {Z.-Y.}\ \bibnamefont {Xue}},\ }\bibinfo {title}
  {High-fidelity geometric gate for silicon-based spin qubits},\ \href
  {\doibase 10.1103/PhysRevA.101.052302} {\bibfield  {journal} {\bibinfo
  {journal} {Phys. Rev. A}\ }\textbf {\bibinfo {volume} {101}},\ \bibinfo
  {pages} {052302} (\bibinfo {year} {2020})}\BibitemShut {NoStop}%
\bibitem [{\citenamefont {Zhou}\ \emph {et~al.}(2021)\citenamefont {Zhou},
  \citenamefont {Li}, \citenamefont {Pan}, \citenamefont {Zhang}, \citenamefont
  {Chen},\ and\ \citenamefont {Xue}}]{PhysRevA.103.032609}%
  \BibitemOpen
  \bibfield  {author} {\bibinfo {author} {\bibfnamefont {J.}~\bibnamefont
  {Zhou}}, \bibinfo {author} {\bibfnamefont {S.}~\bibnamefont {Li}}, \bibinfo
  {author} {\bibfnamefont {G.-Z.}\ \bibnamefont {Pan}}, \bibinfo {author}
  {\bibfnamefont {G.}~\bibnamefont {Zhang}}, \bibinfo {author} {\bibfnamefont
  {T.}~\bibnamefont {Chen}}, \ and\ \bibinfo {author} {\bibfnamefont {Z.-Y.}\
  \bibnamefont {Xue}},\ }\bibinfo {title} {Nonadiabatic geometric quantum gates
  that are insensitive to qubit-frequency drifts},\ \href {\doibase
  10.1103/PhysRevA.103.032609} {\bibfield  {journal} {\bibinfo  {journal}
  {Phys. Rev. A}\ }\textbf {\bibinfo {volume} {103}},\ \bibinfo {pages}
  {032609} (\bibinfo {year} {2021})}\BibitemShut {NoStop}%
\bibitem [{\citenamefont {Xu}\ \emph {et~al.}(2020)\citenamefont {Xu},
  \citenamefont {Hua}, \citenamefont {Chen}, \citenamefont {Pan}, \citenamefont
  {Li}, \citenamefont {Han}, \citenamefont {Cai}, \citenamefont {Ma},
  \citenamefont {Wang}, \citenamefont {Song}, \citenamefont {Xue},\ and\
  \citenamefont {Sun}}]{PhysRevLett.124.230503}%
  \BibitemOpen
  \bibfield  {author} {\bibinfo {author} {\bibfnamefont {Y.}~\bibnamefont
  {Xu}}, \bibinfo {author} {\bibfnamefont {Z.}~\bibnamefont {Hua}}, \bibinfo
  {author} {\bibfnamefont {T.}~\bibnamefont {Chen}}, \bibinfo {author}
  {\bibfnamefont {X.}~\bibnamefont {Pan}}, \bibinfo {author} {\bibfnamefont
  {X.}~\bibnamefont {Li}}, \bibinfo {author} {\bibfnamefont {J.}~\bibnamefont
  {Han}}, \bibinfo {author} {\bibfnamefont {W.}~\bibnamefont {Cai}}, \bibinfo
  {author} {\bibfnamefont {Y.}~\bibnamefont {Ma}}, \bibinfo {author}
  {\bibfnamefont {H.}~\bibnamefont {Wang}}, \bibinfo {author} {\bibfnamefont
  {Y.~P.}\ \bibnamefont {Song}}, \bibinfo {author} {\bibfnamefont {Z.-Y.}\
  \bibnamefont {Xue}}, \ and\ \bibinfo {author} {\bibfnamefont
  {L.}~\bibnamefont {Sun}},\ }\bibinfo {title} {Experimental Implementation of
  Universal Nonadiabatic Geometric Quantum Gates in a Superconducting
  Circuit},\ \href {\doibase 10.1103/PhysRevLett.124.230503} {\bibfield
  {journal} {\bibinfo  {journal} {Phys. Rev. Lett.}\ }\textbf {\bibinfo
  {volume} {124}},\ \bibinfo {pages} {230503} (\bibinfo {year}
  {2020})}\BibitemShut {NoStop}%
\bibitem [{\citenamefont {Zhao}\ \emph {et~al.}(2021)\citenamefont {Zhao},
  \citenamefont {Dong}, \citenamefont {Zhang}, \citenamefont {Guo},
  \citenamefont {Tong},\ and\ \citenamefont {Yin}}]{zhao2021experimental}%
  \BibitemOpen
  \bibfield  {author} {\bibinfo {author} {\bibfnamefont {P.}~\bibnamefont
  {Zhao}}, \bibinfo {author} {\bibfnamefont {Z.}~\bibnamefont {Dong}}, \bibinfo
  {author} {\bibfnamefont {Z.}~\bibnamefont {Zhang}}, \bibinfo {author}
  {\bibfnamefont {G.}~\bibnamefont {Guo}}, \bibinfo {author} {\bibfnamefont
  {D.}~\bibnamefont {Tong}}, \ and\ \bibinfo {author} {\bibfnamefont
  {Y.}~\bibnamefont {Yin}},\ }\bibinfo {title} {Experimental realization of
  nonadiabatic geometric gates with a superconducting Xmon qubit},\ \href@noop
  {} {\bibfield  {journal} {\bibinfo  {journal} {Sci. China Phys. Mech.
  Astron.}\ }\textbf {\bibinfo {volume} {64}},\ \bibinfo {pages} {250362}
  (\bibinfo {year} {2021})}\BibitemShut {NoStop}%
\bibitem [{\citenamefont {Zhou}\ \emph {et~al.}(2025)\citenamefont {Zhou} \emph
  {et~al.}}]{zhou2025experimental}%
  \BibitemOpen
  \bibfield  {author} {\bibinfo {author} {\bibfnamefont {Y.}~\bibnamefont
  {Zhou}} \emph {et~al.},\ }\bibinfo {title} {High-fidelity geometric quantum
  gates exceeding 99.9\% in germanium quantum dots},\ \href@noop {} {\bibfield
  {journal} {\bibinfo  {journal} {Nat. commun.}\ }\textbf {\bibinfo {volume}
  {16}},\ \bibinfo {pages} {7953} (\bibinfo {year} {2025})}\BibitemShut
  {NoStop}%
\bibitem [{\citenamefont {Fang}\ \emph {et~al.}(2024)\citenamefont {Fang},
  \citenamefont {Xu}, \citenamefont {Chen}, \citenamefont {Wei},\ and\
  \citenamefont {Zhang}}]{PhysRevA.109.042615}%
  \BibitemOpen
  \bibfield  {author} {\bibinfo {author} {\bibfnamefont {Z.-Y.}\ \bibnamefont
  {Fang}}, \bibinfo {author} {\bibfnamefont {H.}~\bibnamefont {Xu}}, \bibinfo
  {author} {\bibfnamefont {T.}~\bibnamefont {Chen}}, \bibinfo {author}
  {\bibfnamefont {K.}~\bibnamefont {Wei}}, \ and\ \bibinfo {author}
  {\bibfnamefont {C.}~\bibnamefont {Zhang}},\ }\bibinfo {title} {Nonadiabatic
  geometric quantum gates by composite pulses based on superconducting
  qubits},\ \href {\doibase 10.1103/PhysRevA.109.042615} {\bibfield  {journal}
  {\bibinfo  {journal} {Phys. Rev. A}\ }\textbf {\bibinfo {volume} {109}},\
  \bibinfo {pages} {042615} (\bibinfo {year} {2024})}\BibitemShut {NoStop}%
\bibitem [{\citenamefont {Ding}\ \emph {et~al.}(2026)\citenamefont {Ding},
  \citenamefont {Liu}, \citenamefont {Zhang}, \citenamefont {Zhou},\ and\
  \citenamefont {Xue}}]{ding2025composite}%
  \BibitemOpen
  \bibfield  {author} {\bibinfo {author} {\bibfnamefont {C.-Y.}\ \bibnamefont
  {Ding}}, \bibinfo {author} {\bibfnamefont {W.-F.}\ \bibnamefont {Liu}},
  \bibinfo {author} {\bibfnamefont {L.-H.}\ \bibnamefont {Zhang}}, \bibinfo
  {author} {\bibfnamefont {J.}~\bibnamefont {Zhou}}, \ and\ \bibinfo {author}
  {\bibfnamefont {Z.-Y.}\ \bibnamefont {Xue}},\ }\bibinfo {title} {Composite
  nonadiabatic geometric quantum gates with optimization on superconducting
  circuits},\ \href@noop {} {\bibfield  {journal} {\bibinfo  {journal} {Front.
  Phys.}\ }\textbf {\bibinfo {volume} {21}},\ \bibinfo {pages} {093203}
  (\bibinfo {year} {2026})}\BibitemShut {NoStop}%
\bibitem [{\citenamefont {Chen}\ and\ \citenamefont
  {Xue}(2020)}]{PhysRevApplied.14.064009}%
  \BibitemOpen
  \bibfield  {author} {\bibinfo {author} {\bibfnamefont {T.}~\bibnamefont
  {Chen}}\ and\ \bibinfo {author} {\bibfnamefont {Z.-Y.}\ \bibnamefont {Xue}},\
  }\bibinfo {title} {High-Fidelity and Robust Geometric Quantum Gates that
  Outperform Dynamical Ones},\ \href {\doibase
  10.1103/PhysRevApplied.14.064009} {\bibfield  {journal} {\bibinfo  {journal}
  {Phys. Rev. Appl.}\ }\textbf {\bibinfo {volume} {14}},\ \bibinfo {pages}
  {064009} (\bibinfo {year} {2020})}\BibitemShut {NoStop}%
\bibitem [{\citenamefont {Li}\ \emph {et~al.}(2023)\citenamefont {Li},
  \citenamefont {Liang},\ and\ \citenamefont {Xue}}]{PhysRevA.108.042617}%
  \BibitemOpen
  \bibfield  {author} {\bibinfo {author} {\bibfnamefont {Z.}~\bibnamefont
  {Li}}, \bibinfo {author} {\bibfnamefont {M.-J.}\ \bibnamefont {Liang}}, \
  and\ \bibinfo {author} {\bibfnamefont {Z.-Y.}\ \bibnamefont {Xue}},\
  }\bibinfo {title} {Time-optimal universal quantum gates on superconducting
  circuits},\ \href {\doibase 10.1103/PhysRevA.108.042617} {\bibfield
  {journal} {\bibinfo  {journal} {Phys. Rev. A}\ }\textbf {\bibinfo {volume}
  {108}},\ \bibinfo {pages} {042617} (\bibinfo {year} {2023})}\BibitemShut
  {NoStop}%
\bibitem [{\citenamefont {Xu}\ and\ \citenamefont
  {Long}(2014)}]{PhysRevA.90.022323}%
  \BibitemOpen
  \bibfield  {author} {\bibinfo {author} {\bibfnamefont {G.}~\bibnamefont
  {Xu}}\ and\ \bibinfo {author} {\bibfnamefont {G.}~\bibnamefont {Long}},\
  }\bibinfo {title} {Protecting geometric gates by dynamical decoupling},\
  \href {\doibase 10.1103/PhysRevA.90.022323} {\bibfield  {journal} {\bibinfo
  {journal} {Phys. Rev. A}\ }\textbf {\bibinfo {volume} {90}},\ \bibinfo
  {pages} {022323} (\bibinfo {year} {2014})}\BibitemShut {NoStop}%
\bibitem [{\citenamefont {Wu}\ and\ \citenamefont
  {Zhao}(2020)}]{PhysRevA.102.032627}%
  \BibitemOpen
  \bibfield  {author} {\bibinfo {author} {\bibfnamefont {X.}~\bibnamefont
  {Wu}}\ and\ \bibinfo {author} {\bibfnamefont {P.~Z.}\ \bibnamefont {Zhao}},\
  }\bibinfo {title} {Universal nonadiabatic geometric gates protected by
  dynamical decoupling},\ \href {\doibase 10.1103/PhysRevA.102.032627}
  {\bibfield  {journal} {\bibinfo  {journal} {Phys. Rev. A}\ }\textbf {\bibinfo
  {volume} {102}},\ \bibinfo {pages} {032627} (\bibinfo {year}
  {2020})}\BibitemShut {NoStop}%
\bibitem [{\citenamefont {Li}\ \emph {et~al.}(2020)\citenamefont {Li},
  \citenamefont {Zhao},\ and\ \citenamefont {Tong}}]{PhysRevResearch.2.023295}%
  \BibitemOpen
  \bibfield  {author} {\bibinfo {author} {\bibfnamefont {K.~Z.}\ \bibnamefont
  {Li}}, \bibinfo {author} {\bibfnamefont {P.~Z.}\ \bibnamefont {Zhao}}, \ and\
  \bibinfo {author} {\bibfnamefont {D.~M.}\ \bibnamefont {Tong}},\ }\bibinfo
  {title} {Approach to realizing nonadiabatic geometric gates with prescribed
  evolution paths},\ \href {\doibase 10.1103/PhysRevResearch.2.023295}
  {\bibfield  {journal} {\bibinfo  {journal} {Phys. Rev. Res.}\ }\textbf
  {\bibinfo {volume} {2}},\ \bibinfo {pages} {023295} (\bibinfo {year}
  {2020})}\BibitemShut {NoStop}%
\bibitem [{\citenamefont {Ding}\ \emph {et~al.}(2021)\citenamefont {Ding},
  \citenamefont {Liang}, \citenamefont {Yu},\ and\ \citenamefont
  {Xue}}]{ding2021nonadiabatic}%
  \BibitemOpen
  \bibfield  {author} {\bibinfo {author} {\bibfnamefont {C.-Y.}\ \bibnamefont
  {Ding}}, \bibinfo {author} {\bibfnamefont {Y.}~\bibnamefont {Liang}},
  \bibinfo {author} {\bibfnamefont {K.-Z.}\ \bibnamefont {Yu}}, \ and\ \bibinfo
  {author} {\bibfnamefont {Z.-Y.}\ \bibnamefont {Xue}},\ }\bibinfo {title}
  {Nonadiabatic geometric quantum computation with shortened path on
  superconducting circuits},\ \href@noop {} {\bibfield  {journal} {\bibinfo
  {journal} {Appl. Phys. Lett.}\ }\textbf {\bibinfo {volume} {119}} (\bibinfo
  {year} {2021})}\BibitemShut {NoStop}%
\bibitem [{\citenamefont {Ding}\ \emph {et~al.}(2022)\citenamefont {Ding},
  \citenamefont {Ji}, \citenamefont {Chen},\ and\ \citenamefont
  {Xue}}]{ding2022path}%
  \BibitemOpen
  \bibfield  {author} {\bibinfo {author} {\bibfnamefont {C.-Y.}\ \bibnamefont
  {Ding}}, \bibinfo {author} {\bibfnamefont {L.-N.}\ \bibnamefont {Ji}},
  \bibinfo {author} {\bibfnamefont {T.}~\bibnamefont {Chen}}, \ and\ \bibinfo
  {author} {\bibfnamefont {Z.-Y.}\ \bibnamefont {Xue}},\ }\bibinfo {title}
  {Path-optimized nonadiabatic geometric quantum computation on superconducting
  qubits},\ \href@noop {} {\bibfield  {journal} {\bibinfo  {journal} {Quantum
  Sci. Technol.}\ }\textbf {\bibinfo {volume} {7}},\ \bibinfo {pages} {015012}
  (\bibinfo {year} {2022})}\BibitemShut {NoStop}%
\bibitem [{\citenamefont {Yang}\ \emph {et~al.}(2023)\citenamefont {Yang},
  \citenamefont {Guo}, \citenamefont {Zhang}, \citenamefont {Du}, \citenamefont
  {Zhang}, \citenamefont {Tao}, \citenamefont {Chen}, \citenamefont {Duan},
  \citenamefont {Jia}, \citenamefont {Kong},\ and\ \citenamefont
  {Guo}}]{PhysRevApplied.19.044076}%
  \BibitemOpen
  \bibfield  {author} {\bibinfo {author} {\bibfnamefont {X.-X.}\ \bibnamefont
  {Yang}}, \bibinfo {author} {\bibfnamefont {L.-L.}\ \bibnamefont {Guo}},
  \bibinfo {author} {\bibfnamefont {H.-F.}\ \bibnamefont {Zhang}}, \bibinfo
  {author} {\bibfnamefont {L.}~\bibnamefont {Du}}, \bibinfo {author}
  {\bibfnamefont {C.}~\bibnamefont {Zhang}}, \bibinfo {author} {\bibfnamefont
  {H.-R.}\ \bibnamefont {Tao}}, \bibinfo {author} {\bibfnamefont
  {Y.}~\bibnamefont {Chen}}, \bibinfo {author} {\bibfnamefont {P.}~\bibnamefont
  {Duan}}, \bibinfo {author} {\bibfnamefont {Z.-L.}\ \bibnamefont {Jia}},
  \bibinfo {author} {\bibfnamefont {W.-C.}\ \bibnamefont {Kong}}, \ and\
  \bibinfo {author} {\bibfnamefont {G.-P.}\ \bibnamefont {Guo}},\ }\bibinfo
  {title} {Experimental Implementation of Short-Path Nonadiabatic Geometric
  Gates in a Superconducting Circuit},\ \href {\doibase
  10.1103/PhysRevApplied.19.044076} {\bibfield  {journal} {\bibinfo  {journal}
  {Phys. Rev. Appl.}\ }\textbf {\bibinfo {volume} {19}},\ \bibinfo {pages}
  {044076} (\bibinfo {year} {2023})}\BibitemShut {NoStop}%
\bibitem [{\citenamefont {Liang}\ and\ \citenamefont
  {Xue}(2024)}]{PhysRevApplied.21.064048}%
  \BibitemOpen
  \bibfield  {author} {\bibinfo {author} {\bibfnamefont {Y.}~\bibnamefont
  {Liang}}\ and\ \bibinfo {author} {\bibfnamefont {Z.-Y.}\ \bibnamefont
  {Xue}},\ }\bibinfo {title} {Nonadiabatic geometric quantum gates with
  on-demand trajectories},\ \href {\doibase 10.1103/PhysRevApplied.21.064048}
  {\bibfield  {journal} {\bibinfo  {journal} {Phys. Rev. Appl.}\ }\textbf
  {\bibinfo {volume} {21}},\ \bibinfo {pages} {064048} (\bibinfo {year}
  {2024})}\BibitemShut {NoStop}%
\bibitem [{\citenamefont {Chen}\ \emph {et~al.}(2024)\citenamefont {Chen},
  \citenamefont {Hu}, \citenamefont {Zhang},\ and\ \citenamefont
  {Xue}}]{PhysRevApplied.22.014060}%
  \BibitemOpen
  \bibfield  {author} {\bibinfo {author} {\bibfnamefont {T.}~\bibnamefont
  {Chen}}, \bibinfo {author} {\bibfnamefont {J.-Q.}\ \bibnamefont {Hu}},
  \bibinfo {author} {\bibfnamefont {C.}~\bibnamefont {Zhang}}, \ and\ \bibinfo
  {author} {\bibfnamefont {Z.-Y.}\ \bibnamefont {Xue}},\ }\bibinfo {title}
  {Universal robust geometric quantum control via geometric trajectory
  correction},\ \href {\doibase 10.1103/PhysRevApplied.22.014060} {\bibfield
  {journal} {\bibinfo  {journal} {Phys. Rev. Appl.}\ }\textbf {\bibinfo
  {volume} {22}},\ \bibinfo {pages} {014060} (\bibinfo {year}
  {2024})}\BibitemShut {NoStop}%
\bibitem [{\citenamefont {Roth}\ \emph {et~al.}(2017)\citenamefont {Roth},
  \citenamefont {Ganzhorn}, \citenamefont {Moll}, \citenamefont {Filipp},
  \citenamefont {Salis},\ and\ \citenamefont {Schmidt}}]{PhysRevA.96.062323}%
  \BibitemOpen
  \bibfield  {author} {\bibinfo {author} {\bibfnamefont {M.}~\bibnamefont
  {Roth}}, \bibinfo {author} {\bibfnamefont {M.}~\bibnamefont {Ganzhorn}},
  \bibinfo {author} {\bibfnamefont {N.}~\bibnamefont {Moll}}, \bibinfo {author}
  {\bibfnamefont {S.}~\bibnamefont {Filipp}}, \bibinfo {author} {\bibfnamefont
  {G.}~\bibnamefont {Salis}}, \ and\ \bibinfo {author} {\bibfnamefont
  {S.}~\bibnamefont {Schmidt}},\ }\bibinfo {title} {Analysis of a
  parametrically driven exchange-type gate and a two-photon excitation gate
  between superconducting qubits},\ \href {\doibase 10.1103/PhysRevA.96.062323}
  {\bibfield  {journal} {\bibinfo  {journal} {Phys. Rev. A}\ }\textbf {\bibinfo
  {volume} {96}},\ \bibinfo {pages} {062323} (\bibinfo {year}
  {2017})}\BibitemShut {NoStop}%
\bibitem [{\citenamefont {Reagor}\ \emph {et~al.}(2018)\citenamefont {Reagor},
  \citenamefont {Osborn}, \citenamefont {Tezak}, \citenamefont {Staley},
  \citenamefont {Prawiroatmodjo}, \citenamefont {Scheer}, \citenamefont
  {Alidoust}, \citenamefont {Sete}, \citenamefont {Didier}, \citenamefont
  {da~Silva} \emph {et~al.}}]{reagor2018demonstration}%
  \BibitemOpen
  \bibfield  {author} {\bibinfo {author} {\bibfnamefont {M.}~\bibnamefont
  {Reagor}}, \bibinfo {author} {\bibfnamefont {C.~B.}\ \bibnamefont {Osborn}},
  \bibinfo {author} {\bibfnamefont {N.}~\bibnamefont {Tezak}}, \bibinfo
  {author} {\bibfnamefont {A.}~\bibnamefont {Staley}}, \bibinfo {author}
  {\bibfnamefont {G.}~\bibnamefont {Prawiroatmodjo}}, \bibinfo {author}
  {\bibfnamefont {M.}~\bibnamefont {Scheer}}, \bibinfo {author} {\bibfnamefont
  {N.}~\bibnamefont {Alidoust}}, \bibinfo {author} {\bibfnamefont {E.~A.}\
  \bibnamefont {Sete}}, \bibinfo {author} {\bibfnamefont {N.}~\bibnamefont
  {Didier}}, \bibinfo {author} {\bibfnamefont {M.~P.}\ \bibnamefont
  {da~Silva}},  \emph {et~al.},\ }\bibinfo {title} {Demonstration of universal
  parametric entangling gates on a multi-qubit lattice},\ \href@noop {}
  {\bibfield  {journal} {\bibinfo  {journal} {Sci. Adv.}\ }\textbf {\bibinfo
  {volume} {4}},\ \bibinfo {pages} {eaao3603} (\bibinfo {year}
  {2018})}\BibitemShut {NoStop}%
\bibitem [{\citenamefont {Caldwell}\ \emph {et~al.}(2018)\citenamefont
  {Caldwell}, \citenamefont {Didier}, \citenamefont {Ryan} \emph
  {et~al.}}]{PhysRevApplied.10.034050}%
  \BibitemOpen
  \bibfield  {author} {\bibinfo {author} {\bibfnamefont {S.~A.}\ \bibnamefont
  {Caldwell}}, \bibinfo {author} {\bibfnamefont {N.}~\bibnamefont {Didier}},
  \bibinfo {author} {\bibfnamefont {C.~A.}\ \bibnamefont {Ryan}},  \emph
  {et~al.},\ }\bibinfo {title} {Parametrically Activated Entangling Gates Using
  Transmon Qubits},\ \href {\doibase 10.1103/PhysRevApplied.10.034050}
  {\bibfield  {journal} {\bibinfo  {journal} {Phys. Rev. Appl.}\ }\textbf
  {\bibinfo {volume} {10}},\ \bibinfo {pages} {034050} (\bibinfo {year}
  {2018})}\BibitemShut {NoStop}%
\bibitem [{\citenamefont {Li}\ \emph {et~al.}(2018)\citenamefont {Li},
  \citenamefont {Ma}, \citenamefont {Han}, \citenamefont {Chen}, \citenamefont
  {Xu}, \citenamefont {Cai}, \citenamefont {Wang}, \citenamefont {Song},
  \citenamefont {Xue}, \citenamefont {Yin},\ and\ \citenamefont
  {Sun}}]{PhysRevApplied.10.054009}%
  \BibitemOpen
  \bibfield  {author} {\bibinfo {author} {\bibfnamefont {X.}~\bibnamefont
  {Li}}, \bibinfo {author} {\bibfnamefont {Y.}~\bibnamefont {Ma}}, \bibinfo
  {author} {\bibfnamefont {J.}~\bibnamefont {Han}}, \bibinfo {author}
  {\bibfnamefont {T.}~\bibnamefont {Chen}}, \bibinfo {author} {\bibfnamefont
  {Y.}~\bibnamefont {Xu}}, \bibinfo {author} {\bibfnamefont {W.}~\bibnamefont
  {Cai}}, \bibinfo {author} {\bibfnamefont {H.}~\bibnamefont {Wang}}, \bibinfo
  {author} {\bibfnamefont {Y.}~\bibnamefont {Song}}, \bibinfo {author}
  {\bibfnamefont {Z.-Y.}\ \bibnamefont {Xue}}, \bibinfo {author} {\bibfnamefont
  {Z.-q.}\ \bibnamefont {Yin}}, \ and\ \bibinfo {author} {\bibfnamefont
  {L.}~\bibnamefont {Sun}},\ }\bibinfo {title} {Perfect Quantum State Transfer
  in a Superconducting Qubit Chain with Parametrically Tunable Couplings},\
  \href {\doibase 10.1103/PhysRevApplied.10.054009} {\bibfield  {journal}
  {\bibinfo  {journal} {Phys. Rev. Appl.}\ }\textbf {\bibinfo {volume} {10}},\
  \bibinfo {pages} {054009} (\bibinfo {year} {2018})}\BibitemShut {NoStop}%
\bibitem [{\citenamefont {Wang}\ \emph {et~al.}(2009)\citenamefont {Wang},
  \citenamefont {Sun},\ and\ \citenamefont {Wang}}]{PhysRevA.79.012105}%
  \BibitemOpen
  \bibfield  {author} {\bibinfo {author} {\bibfnamefont {X.}~\bibnamefont
  {Wang}}, \bibinfo {author} {\bibfnamefont {Z.}~\bibnamefont {Sun}}, \ and\
  \bibinfo {author} {\bibfnamefont {Z.~D.}\ \bibnamefont {Wang}},\ }\bibinfo
  {title} {Operator fidelity susceptibility: An indicator of quantum
  criticality},\ \href {\doibase 10.1103/PhysRevA.79.012105} {\bibfield
  {journal} {\bibinfo  {journal} {Phys. Rev. A}\ }\textbf {\bibinfo {volume}
  {79}},\ \bibinfo {pages} {012105} (\bibinfo {year} {2009})}\BibitemShut
  {NoStop}%
\bibitem [{\citenamefont {Koch}\ \emph {et~al.}(2007)\citenamefont {Koch},
  \citenamefont {Yu}, \citenamefont {Gambetta}, \citenamefont {Houck},
  \citenamefont {Schuster}, \citenamefont {Majer}, \citenamefont {Blais},
  \citenamefont {Devoret}, \citenamefont {Girvin},\ and\ \citenamefont
  {Schoelkopf}}]{PhysRevA.76.042319}%
  \BibitemOpen
  \bibfield  {author} {\bibinfo {author} {\bibfnamefont {J.}~\bibnamefont
  {Koch}}, \bibinfo {author} {\bibfnamefont {T.~M.}\ \bibnamefont {Yu}},
  \bibinfo {author} {\bibfnamefont {J.}~\bibnamefont {Gambetta}}, \bibinfo
  {author} {\bibfnamefont {A.~A.}\ \bibnamefont {Houck}}, \bibinfo {author}
  {\bibfnamefont {D.~I.}\ \bibnamefont {Schuster}}, \bibinfo {author}
  {\bibfnamefont {J.}~\bibnamefont {Majer}}, \bibinfo {author} {\bibfnamefont
  {A.}~\bibnamefont {Blais}}, \bibinfo {author} {\bibfnamefont {M.~H.}\
  \bibnamefont {Devoret}}, \bibinfo {author} {\bibfnamefont {S.~M.}\
  \bibnamefont {Girvin}}, \ and\ \bibinfo {author} {\bibfnamefont {R.~J.}\
  \bibnamefont {Schoelkopf}},\ }\bibinfo {title} {Charge-insensitive qubit
  design derived from the Cooper pair box},\ \href {\doibase
  10.1103/PhysRevA.76.042319} {\bibfield  {journal} {\bibinfo  {journal} {Phys.
  Rev. A}\ }\textbf {\bibinfo {volume} {76}},\ \bibinfo {pages} {042319}
  (\bibinfo {year} {2007})}\BibitemShut {NoStop}%
\bibitem [{\citenamefont {You}\ \emph {et~al.}(2007)\citenamefont {You},
  \citenamefont {Hu}, \citenamefont {Ashhab},\ and\ \citenamefont
  {Nori}}]{PhysRevB.75.140515}%
  \BibitemOpen
  \bibfield  {author} {\bibinfo {author} {\bibfnamefont {J.~Q.}\ \bibnamefont
  {You}}, \bibinfo {author} {\bibfnamefont {X.}~\bibnamefont {Hu}}, \bibinfo
  {author} {\bibfnamefont {S.}~\bibnamefont {Ashhab}}, \ and\ \bibinfo {author}
  {\bibfnamefont {F.}~\bibnamefont {Nori}},\ }\bibinfo {title} {Low-decoherence
  flux qubit},\ \href {\doibase 10.1103/PhysRevB.75.140515} {\bibfield
  {journal} {\bibinfo  {journal} {Phys. Rev. B}\ }\textbf {\bibinfo {volume}
  {75}},\ \bibinfo {pages} {140515(R)} (\bibinfo {year} {2007})}\BibitemShut
  {NoStop}%
\bibitem [{\citenamefont {Krantz}\ \emph {et~al.}(2019)\citenamefont {Krantz},
  \citenamefont {Kjaergaard}, \citenamefont {Yan}, \citenamefont {Orlando},
  \citenamefont {Gustavsson},\ and\ \citenamefont
  {Oliver}}]{krantz2019quantum}%
  \BibitemOpen
  \bibfield  {author} {\bibinfo {author} {\bibfnamefont {P.}~\bibnamefont
  {Krantz}}, \bibinfo {author} {\bibfnamefont {M.}~\bibnamefont {Kjaergaard}},
  \bibinfo {author} {\bibfnamefont {F.}~\bibnamefont {Yan}}, \bibinfo {author}
  {\bibfnamefont {T.~P.}\ \bibnamefont {Orlando}}, \bibinfo {author}
  {\bibfnamefont {S.}~\bibnamefont {Gustavsson}}, \ and\ \bibinfo {author}
  {\bibfnamefont {W.~D.}\ \bibnamefont {Oliver}},\ }\bibinfo {title} {A quantum
  engineer's guide to superconducting qubits},\ \href@noop {} {\bibfield
  {journal} {\bibinfo  {journal} {Appl. Phys. Rev.}\ }\textbf {\bibinfo
  {volume} {6}},\ \bibinfo {pages} {021318} (\bibinfo {year}
  {2019})}\BibitemShut {NoStop}%
\bibitem [{\citenamefont {Motzoi}\ \emph {et~al.}(2009)\citenamefont {Motzoi},
  \citenamefont {Gambetta}, \citenamefont {Rebentrost},\ and\ \citenamefont
  {Wilhelm}}]{PhysRevLett.103.110501}%
  \BibitemOpen
  \bibfield  {author} {\bibinfo {author} {\bibfnamefont {F.}~\bibnamefont
  {Motzoi}}, \bibinfo {author} {\bibfnamefont {J.~M.}\ \bibnamefont
  {Gambetta}}, \bibinfo {author} {\bibfnamefont {P.}~\bibnamefont
  {Rebentrost}}, \ and\ \bibinfo {author} {\bibfnamefont {F.~K.}\ \bibnamefont
  {Wilhelm}},\ }\bibinfo {title} {Simple Pulses for Elimination of Leakage in
  Weakly Nonlinear Qubits},\ \href {\doibase 10.1103/PhysRevLett.103.110501}
  {\bibfield  {journal} {\bibinfo  {journal} {Phys. Rev. Lett.}\ }\textbf
  {\bibinfo {volume} {103}},\ \bibinfo {pages} {110501} (\bibinfo {year}
  {2009})}\BibitemShut {NoStop}%
\bibitem [{\citenamefont {Chen}\ \emph {et~al.}(2016)\citenamefont {Chen},
  \citenamefont {Kelly}, \citenamefont {Quintana}, \citenamefont {Barends},
  \citenamefont {Campbell}, \citenamefont {Chen}, \citenamefont {Chiaro},
  \citenamefont {Dunsworth}, \citenamefont {Fowler}, \citenamefont {Lucero}
  \emph {et~al.}}]{PhysRevLett.116.020501}%
  \BibitemOpen
  \bibfield  {author} {\bibinfo {author} {\bibfnamefont {Z.}~\bibnamefont
  {Chen}}, \bibinfo {author} {\bibfnamefont {J.}~\bibnamefont {Kelly}},
  \bibinfo {author} {\bibfnamefont {C.}~\bibnamefont {Quintana}}, \bibinfo
  {author} {\bibfnamefont {R.}~\bibnamefont {Barends}}, \bibinfo {author}
  {\bibfnamefont {B.}~\bibnamefont {Campbell}}, \bibinfo {author}
  {\bibfnamefont {Y.}~\bibnamefont {Chen}}, \bibinfo {author} {\bibfnamefont
  {B.}~\bibnamefont {Chiaro}}, \bibinfo {author} {\bibfnamefont
  {A.}~\bibnamefont {Dunsworth}}, \bibinfo {author} {\bibfnamefont {A.~G.}\
  \bibnamefont {Fowler}}, \bibinfo {author} {\bibfnamefont {E.}~\bibnamefont
  {Lucero}},  \emph {et~al.},\ }\bibinfo {title} {Measuring and Suppressing
  Quantum State Leakage in a Superconducting Qubit},\ \href {\doibase
  10.1103/PhysRevLett.116.020501} {\bibfield  {journal} {\bibinfo  {journal}
  {Phys. Rev. Lett.}\ }\textbf {\bibinfo {volume} {116}},\ \bibinfo {pages}
  {020501} (\bibinfo {year} {2016})}\BibitemShut {NoStop}%
\bibitem [{\citenamefont {Lindblad}(1976)}]{lindblad1976generators}%
  \BibitemOpen
  \bibfield  {author} {\bibinfo {author} {\bibfnamefont {G.}~\bibnamefont
  {Lindblad}},\ }\bibinfo {title} {On the generators of quantum dynamical
  semigroups},\ \href@noop {} {\bibfield  {journal} {\bibinfo  {journal}
  {Commun. Math. Phys.}\ }\textbf {\bibinfo {volume} {48}},\ \bibinfo {pages}
  {119} (\bibinfo {year} {1976})}\BibitemShut {NoStop}%
\bibitem [{\citenamefont {Poyatos}\ \emph {et~al.}(1997)\citenamefont
  {Poyatos}, \citenamefont {Cirac},\ and\ \citenamefont
  {Zoller}}]{PhysRevLett.78.390}%
  \BibitemOpen
  \bibfield  {author} {\bibinfo {author} {\bibfnamefont {J.~F.}\ \bibnamefont
  {Poyatos}}, \bibinfo {author} {\bibfnamefont {J.~I.}\ \bibnamefont {Cirac}},
  \ and\ \bibinfo {author} {\bibfnamefont {P.}~\bibnamefont {Zoller}},\
  }\bibinfo {title} {Complete Characterization of a Quantum Process: The
  Two-Bit Quantum Gate},\ \href {\doibase 10.1103/PhysRevLett.78.390}
  {\bibfield  {journal} {\bibinfo  {journal} {Phys. Rev. Lett.}\ }\textbf
  {\bibinfo {volume} {78}},\ \bibinfo {pages} {390} (\bibinfo {year}
  {1997})}\BibitemShut {NoStop}%
\bibitem [{\citenamefont {Kjaergaard}\ \emph {et~al.}(2020)\citenamefont
  {Kjaergaard}, \citenamefont {Schwartz}, \citenamefont {Braum{\"u}ller},
  \citenamefont {Krantz}, \citenamefont {Wang}, \citenamefont {Gustavsson},\
  and\ \citenamefont {Oliver}}]{kjaergaard2020superconducting}%
  \BibitemOpen
  \bibfield  {author} {\bibinfo {author} {\bibfnamefont {M.}~\bibnamefont
  {Kjaergaard}}, \bibinfo {author} {\bibfnamefont {M.~E.}\ \bibnamefont
  {Schwartz}}, \bibinfo {author} {\bibfnamefont {J.}~\bibnamefont
  {Braum{\"u}ller}}, \bibinfo {author} {\bibfnamefont {P.}~\bibnamefont
  {Krantz}}, \bibinfo {author} {\bibfnamefont {J.~I.-J.}\ \bibnamefont {Wang}},
  \bibinfo {author} {\bibfnamefont {S.}~\bibnamefont {Gustavsson}}, \ and\
  \bibinfo {author} {\bibfnamefont {W.~D.}\ \bibnamefont {Oliver}},\ }\bibinfo
  {title} {Superconducting qubits: Current state of play},\ \href@noop {}
  {\bibfield  {journal} {\bibinfo  {journal} {Annu. Rev. Condens. Matter
  Phys.}\ }\textbf {\bibinfo {volume} {11}},\ \bibinfo {pages} {369} (\bibinfo
  {year} {2020})}\BibitemShut {NoStop}%
\bibitem [{\citenamefont {Huang}\ \emph {et~al.}(2020)\citenamefont {Huang},
  \citenamefont {Wu}, \citenamefont {Fan},\ and\ \citenamefont
  {Zhu}}]{huang2020superconducting}%
  \BibitemOpen
  \bibfield  {author} {\bibinfo {author} {\bibfnamefont {H.-L.}\ \bibnamefont
  {Huang}}, \bibinfo {author} {\bibfnamefont {D.}~\bibnamefont {Wu}}, \bibinfo
  {author} {\bibfnamefont {D.}~\bibnamefont {Fan}}, \ and\ \bibinfo {author}
  {\bibfnamefont {X.}~\bibnamefont {Zhu}},\ }\bibinfo {title} {Superconducting
  quantum computing: a review},\ \href@noop {} {\bibfield  {journal} {\bibinfo
  {journal} {Sci. China Inf. Sci.}\ }\textbf {\bibinfo {volume} {63}},\
  \bibinfo {pages} {180501} (\bibinfo {year} {2020})}\BibitemShut {NoStop}%
\bibitem [{\citenamefont {Stehlik}\ \emph {et~al.}(2021)\citenamefont
  {Stehlik}, \citenamefont {Zajac}, \citenamefont {Underwood}, \citenamefont
  {Phung}, \citenamefont {Blair}, \citenamefont {Carnevale}, \citenamefont
  {Klaus}, \citenamefont {Keefe}, \citenamefont {Carniol}, \citenamefont
  {Kumph}, \citenamefont {Steffen},\ and\ \citenamefont
  {Dial}}]{PhysRevLett.127.080505}%
  \BibitemOpen
  \bibfield  {author} {\bibinfo {author} {\bibfnamefont {J.}~\bibnamefont
  {Stehlik}}, \bibinfo {author} {\bibfnamefont {D.~M.}\ \bibnamefont {Zajac}},
  \bibinfo {author} {\bibfnamefont {D.~L.}\ \bibnamefont {Underwood}}, \bibinfo
  {author} {\bibfnamefont {T.}~\bibnamefont {Phung}}, \bibinfo {author}
  {\bibfnamefont {J.}~\bibnamefont {Blair}}, \bibinfo {author} {\bibfnamefont
  {S.}~\bibnamefont {Carnevale}}, \bibinfo {author} {\bibfnamefont
  {D.}~\bibnamefont {Klaus}}, \bibinfo {author} {\bibfnamefont {G.~A.}\
  \bibnamefont {Keefe}}, \bibinfo {author} {\bibfnamefont {A.}~\bibnamefont
  {Carniol}}, \bibinfo {author} {\bibfnamefont {M.}~\bibnamefont {Kumph}},
  \bibinfo {author} {\bibfnamefont {M.}~\bibnamefont {Steffen}}, \ and\
  \bibinfo {author} {\bibfnamefont {O.~E.}\ \bibnamefont {Dial}},\ }\bibinfo
  {title} {Tunable Coupling Architecture for Fixed-Frequency Transmon
  Superconducting Qubits},\ \href {\doibase 10.1103/PhysRevLett.127.080505}
  {\bibfield  {journal} {\bibinfo  {journal} {Phys. Rev. Lett.}\ }\textbf
  {\bibinfo {volume} {127}},\ \bibinfo {pages} {080505} (\bibinfo {year}
  {2021})}\BibitemShut {NoStop}%
\bibitem [{\citenamefont {Ni}\ \emph {et~al.}(2022)\citenamefont {Ni},
  \citenamefont {Li}, \citenamefont {Zhang}, \citenamefont {Chu}, \citenamefont
  {Niu}, \citenamefont {Yan}, \citenamefont {Deng}, \citenamefont {Hu},
  \citenamefont {Li}, \citenamefont {Zhong}, \citenamefont {Liu}, \citenamefont
  {Yan}, \citenamefont {Xu},\ and\ \citenamefont
  {Yu}}]{PhysRevLett.129.040502}%
  \BibitemOpen
  \bibfield  {author} {\bibinfo {author} {\bibfnamefont {Z.}~\bibnamefont
  {Ni}}, \bibinfo {author} {\bibfnamefont {S.}~\bibnamefont {Li}}, \bibinfo
  {author} {\bibfnamefont {L.}~\bibnamefont {Zhang}}, \bibinfo {author}
  {\bibfnamefont {J.}~\bibnamefont {Chu}}, \bibinfo {author} {\bibfnamefont
  {J.}~\bibnamefont {Niu}}, \bibinfo {author} {\bibfnamefont {T.}~\bibnamefont
  {Yan}}, \bibinfo {author} {\bibfnamefont {X.}~\bibnamefont {Deng}}, \bibinfo
  {author} {\bibfnamefont {L.}~\bibnamefont {Hu}}, \bibinfo {author}
  {\bibfnamefont {J.}~\bibnamefont {Li}}, \bibinfo {author} {\bibfnamefont
  {Y.}~\bibnamefont {Zhong}}, \bibinfo {author} {\bibfnamefont
  {S.}~\bibnamefont {Liu}}, \bibinfo {author} {\bibfnamefont {F.}~\bibnamefont
  {Yan}}, \bibinfo {author} {\bibfnamefont {Y.}~\bibnamefont {Xu}}, \ and\
  \bibinfo {author} {\bibfnamefont {D.}~\bibnamefont {Yu}},\ }\bibinfo {title}
  {Scalable Method for Eliminating Residual $ZZ$ Interaction between
  Superconducting Qubits},\ \href {\doibase 10.1103/PhysRevLett.129.040502}
  {\bibfield  {journal} {\bibinfo  {journal} {Phys. Rev. Lett.}\ }\textbf
  {\bibinfo {volume} {129}},\ \bibinfo {pages} {040502} (\bibinfo {year}
  {2022})}\BibitemShut {NoStop}%
\end{thebibliography}
%

\end{document}